\documentclass[useAMS,usenatbib]{mn2e}

\makeatletter
\def\thebibliography#1{\section*{REFERENCES}
 \addcontentsline{toc}{section}{REFERENCES}
 \list{}{\usecounter{dummy}%
         \labelwidth 0pt
         \leftmargin 1.5em
         \itemsep 0pt
         \itemindent-\leftmargin}
 \reset@font\small
 \parindent 0pt
 \parskip 0pt plus .1pt\relax
 \def\newblock{\hskip .11em plus .33em minus .07em}
 \sloppy\clubpenalty4000\widowpenalty4000
 \sfcode`\.=1000\relax
}
\def\here{\global\advance\count\@currbox -1}
\makeatother

\usepackage[T1]{fontenc}
\usepackage{ae,aecompl}

\usepackage{amsmath}% Advanced maths commands
\usepackage{color}
\usepackage{graphicx}\graphicspath{{plots/}}
\usepackage{epstopdf}
\usepackage{balance}

\newcommand{\eprint}[2][]{{\tt\if!#1!#2\else#1:#2\fi}}
\newcommand{\Like}{\mathcal{L}}
\usepackage[latin1]{inputenc}
\newcommand{\ds}{\hphantom{0}} % digit space
\newcommand{\ps}{\hphantom{.}} % period space
\DeclareInputText{183}{\ds}    % middle dot in latin1 encoding
\DeclareInputText{172}{\ps}    % not sign in latin1 encoding

\DeclareMathSymbol{:}{\mathord}{operators}{"3A}
\usepackage{amssymb}
\usepackage{tabularx}
\newcommand{\topline}{\hline\\[-7pt]}
\newcommand{\midline}{\\[3pt]\hline\\[-7pt]}
\newcommand{\dblline}{\\[3pt]\hline\hline\\[-7pt]}
\newcommand{\botline}{\\[3pt]\hline}

\title[GAMA: The $z \sim 0$ field galaxy luminosity function to $L \sim 10^6~ \mathrm{L}_\odot$]{Galaxy And Mass Assembly (GAMA): $\mathbf{z \sim 0}$ Galaxy Luminosity Function down to $\mathbf{L \sim 10^{6}~L_\odot}$ via Clustering Based Redshift Inference}

\author[G.\ S.\ Karademir et al.]{Geray S.\ Karademir$^{1}$\thanks{E-mail: gkarademir@swin.edu.au},
Edward N.\ Taylor$^{1}$, 
Chris Blake$^{1}$,
Ivan K.\ Baldry$^2$, \newauthor
Sabine Bellstedt$^3$,
Maciej Bilicki$^4$,
Michael J.\ I.\ Brown$^5$, 
Michelle E.\ Cluver$^1$, \newauthor
Simon P.\ Driver$^3$,
Hendrik Hildebrandt$^6$,
Benne W.\ Holwerda$^7$,
Andrew M.\ Hopkins$^8$, \newauthor
Jonathan Loveday$^9$,
Steven Phillipps$^{10}$,
Angus H. Wright$^{6}$
\vspace{6pt}\\
$^{1}$ Centre for Astrophysics \& Supercomputing, Swinburne University of Technology, PO Box 218, Hawthorn, VIC 3122, Australia;
\\
$^{2}$ Astrophysics Research Institute, Liverpool John Moores University, IC2, Liverpool Science Park, 146 Brownlow Hill, Liverpool, L3 5RF; 
\\
$^3$ ICRAR, The University of Western Australia, 35 Stirling Highway, Crawley, WA 6009, Australia; 
\\
$^4$ Center for Theoretical Physics PAS, al.\ Lotnik\'ow 32/46, 02-668 Warsaw, Poland; 
\\
$^5$ School of Physics and Astronomy, Monash University, Clayton, VIC 3800, Australia; 
\\
$^6$ Ruhr University Bochum, Faculty of Physics and Astronomy, Astronomical Institute (AIRUB), German Centre for Cosmological Lensing, \\
44780 Bochum, Germany;
\\
$^7$ Department of Physics and Astronomy, 102 Natural Science Building, University of Louisville, Louisville KY 40292, USA;
\\
$^8$ Australian Astronomical Optics, Macquarie University, 105 Delhi Rd, North Ryde, NSW 2113, Australia;
\\
$^9$ Astronomy Centre, University of Sussex, Falmer, Brighton, BN1 9QH, UK ;
\\
$^{10}$ Astrophysics Group, School of Physics, University of Bristol, Bristol BS8 1TL, UK
\\
}
\date{Accepted 2021 November 2. Received 2021 October 31; in original form 2021 September 8} % These dates will be filled out by the publisher

\pubyear{2021} % Enter the current year, for the copyright statements etc.

\begin{document}

\label{firstpage}
\pagerange{\pageref{firstpage}--\pageref{lastpage}}
\maketitle

\begin{abstract}
In this study we present a new experimental design using clustering-based redshift inference to measure the evolving galaxy luminosity function (GLF) spanning 5.5 decades from $L \sim 10^{11.5}$ to $ 10^6 ~ \mathrm{L}_\odot$. We use data from the Galaxy And Mass Assembly (GAMA) survey and the Kilo-Degree Survey (KiDS). We derive redshift distributions in bins of apparent magnitude to the limits of the GAMA-KiDS photometric catalogue: $m_r \lesssim 23$; more than a decade in luminosity beyond the limits of the GAMA spectroscopic redshift sample via clustering-based redshift inference. This technique uses spatial cross-correlation statistics for a reference set with known redshifts (in our case, the main GAMA sample) to derive the redshift distribution for the target ensemble. For the calibration of the redshift distribution we use a simple parametrisation with an adaptive normalisation factor over the interval $0.005 < z < 0.48$ to derive the clustering redshift results. We find that the GLF has a relatively constant power-law slope $\alpha \approx -1.2$ for $-17 \lesssim M_r \lesssim -13$, and then appears to steepen sharply for $-13 \lesssim M_r \lesssim -10$. This upturn appears to be where Globular Clusters (GCs) take over to dominate the source counts as a function of luminosity. Thus we have mapped the GLF across the full range of the $z \sim 0$ field galaxy population from the most luminous galaxies down to the GC scale.
\end{abstract}


%%%%%%%%%%%%%%%%%%%%%%%%%%%%%%%%%%%%%%%%%%%%%%%%%%
%%%%%%%%%%%%%%%%% BODY OF PAPER %%%%%%%%%%%%%%%%%%
\section{Introduction}
\label{sec:Intro}
The galaxy luminosity function (GLF) is a basic descriptor of the galaxy population and its evolution though the history of the Universe. GLF measurements \citep[e.g.][]{Sandage1985, Driver1996, Trentham2002} play a key role in calibrating and validating theoretical models of galaxy formation and evolution. For example, energetic feedback by Active Galactic Nuclei (AGN) \citet{Croton2006, Bower2012} has been invoked to explain the exponential drop-off at the bright end of the GLF. At the faint end, the slope of the GLF is usually understood to be determined by the efficiency of gas accretion onto low mass halos \citep[e.g.][]{White1978, Kauffmann1993, Cole1994} and by self-regulated star formation \citep[e.g.\ through supernova feedback;][]{Dekel1986}.

One approach to measuring the GLF down to very faint luminosities has been to target particular structures or environments, including the Local Group \citep[e.g.][]{Trentham2005, Koposov2008}, for selected groups \citep[e.g.][]{Trentham2002, Chiboucas2009, Mao2021}, and in clusters \citep[e.g.][]{Driver1994, Popesso2005}. In contrast to the Local Group the GLF of the Coma Cluster \citep[][]{Yamanoi2012} and multiple Hickson Compact Groups \citep[][]{Yamanoi2020} show a significant upturn of the GLF at $M_r>-12$. \cite{Yamanoi2012} argue that, in clusters, the faint end of the GLF consists of galaxy populations with different origins and that the contribution of Globular Clusters (GC) has to be considered as unresolved low-luminosity galaxies whose angular sizes are similar to the seeing size cannot be distinguished from bright GCs. 

Obtaining robust measurements of the field (i.e. cosmic average) GLF at very low luminosities \citep[e.g.][]{Zucca1997, Loveday1997, Marzke1998, Blanton2005} remains an observational challenge, as it requires very deep data (to probe the faintest luminosities) over very wide areas (to probe significant volumes at low redshifts), as well as good redshift information (to map observed to intrinsic properties, and to distinguish nearby and distant galaxy populations). Early attempts based on spectroscopic redshift surveys suffered from strong surface-brightness (SB) selection effects \citep[e.g.][]{Phillipps1986, McGaugh1996, Cross2002}. While the first measurement of the impact of low surface brightness galaxies was performed by \cite{Sprayberry1997}, \cite{Cross2001} showed that the bias in surface brightness can lead to an underestimation of the GLF, and therefore the luminosity density, by $\sim35\%$. Most recently, the Galaxy And Mass Assembly (GAMA) survey has obtained spectroscopic redshifts with near total completeness for $m_r < 19.8$ over $\sim 220$ sq.\ deg.,\ and has measured the GLF down to $10^{7.5} M_\odot$ with minimal corrections required to account for SB selection effects \citep{Loveday2015, Wright2017}. 

The primary aim of this study is to measure the field GLF down to the faintest possible limits. We do this through a process we call {\em clustering redshift inference}, or cluster-$z$s. This process exploits the fact that galaxies are strongly clustered (rather than randomly distributed) to derive redshift information for our target sample, using only their observed positions on the sky.

That galaxies are strongly clustered, both in real space and projected on the sky, is an essential fact of cosmology \citep[e.g.][]{Peebles1980, Cole2005, Eisenstein2005}. The idea of using angular cross-correlations to trace physical correlations has been used for a few decades \citep[e.g.][]{Seldner1979,Phillipps1985,Phillipps1987,Loveday1997}. The approach to clustering redshift inference was described in greater detail by \cite{Schneider2006,Newman2008,Matthews2010} and \cite{Matthews2012} with a more generalized formalisation presented by \cite{Schmidt2013} and \cite{Menard2013} including validation with numerical simulations. These techniques have been applied to observations as well as simulations by several studies \citep[e.g.][]{McQuinn2013, Rahman2015, Choi2016, Rahman2016a, Rahman2016b, Scottez2016, Johnson2017, Busch2020}. By testing multiple clustering-based methods, \cite{Gatti2018} showed that the systematic error induced by neglecting the redshift evolution of the galaxy bias is the main systematic error associated with this method.

The measurement of luminosity functions from clustering-based redshifts for mock galaxy samples is presented by \cite{vanDaalen2018} and \cite{Bates2019} has used clustering-based redshifts to map the $0.2 < z < 0.8$ evolution of the GLF in small bins of color and magnitude to $m_i < 21$, and used the results to determine redshift-dependent incompleteness corrections for the BOSS survey \citep[][]{BOSS2013}. In this study we aim to probe for the faint end of the $z\sim 0 $ GLF.

Our objective is to use clustering-based redshift inference to measure the $z\sim 0 $ GLF down to the faintest possible limits, beyond the reach of spectroscopic and photometric redshift surveys. The rest of this paper is structured as follows. In Sec. \ref{sec:data} we describe the imaging, photometry, and spectroscopic redshift catalogues that we use for our cluster-$z$ analysis and GLF measurements. The methodology to derive and normalise the clustering-based redshift estimates of a data set with unknown redshift information is described in Sec. \ref{sec:cluster-$z$s}, including verification/validation of our cluster-$z$ results in Sec.\ \ref{sec:cluster-$z$s-validation}. Sec. \ref{sec:fitting} describes our descriptive model for the evolving GLF, which is a critical step for normalising the cluster-$z$ results. In Sec. \ref{sec:results} we present the results of our study as well as the measured $z < 0.1$ luminosity functions to $M_r<-10$. Finally in Sec. \ref{sec:Discussion} and Sec. \ref{sec:Summary} we discuss and summarize the results of our study. Throughout our paper we use a flat $\Lambda$CDM cosmology with $\Omega_M=0.3$, $\Omega_\Lambda=0.7$ and a Hubble parameter $H_0 = 100 \ h$ km Mpc$^{-1}$ s$^{-1}$ where $h = 0.7$. All photometry has been corrected for foreground Galactic extinction using the Planck EBV map \citep{Planck2013}.

\section{Data and sample selection}
\label{sec:data}
The data requirements for our study are as follows. Firstly, we rely on high quality photometry from deep optical imaging to map the apparent fluxes of the evolving galaxy population. We use positions and total $r$-band magnitudes from a GAMA reanalysis of VST imaging from the KiDS survey, described in Sec. \ref{sec:photometry}, to define our target sample as described in Sec. \ref{sec:selection}. The relevant selection effects limiting our analysis are discussed in Sec. \ref{sec:limits}.

\subsection{Positions and Total Photometry for the Target Sample}
\label{sec:photometry}
The Kilo-Degree Survey \citep[KiDS;][]{KiDSKuijken2019} is a deep, wide-field optical imaging survey using ESO's VLT Survey Telescope (VST) with the primary motivation of weak lensing science \citep[e.g.][]{Hildebrandt2016}. KiDS has obtained $ugri$ imaging with sub-arcsecond seeing and nearly uniform depth over $\sim 1350$ square degrees. For the $r$-band data, which is the focus of this study, the median seeing is $< 0.6^{\prime\prime}$ FHWM and the 5$\sigma$ point source magnitude limit is 25.2 mag. The fourth KiDS Data Release \citep{KiDSKuijken2019} made public over 1000 square degrees of imaging, including 4 GAMA survey fields. In our study we focus on the three equatorial 60 square degree fields of GAMA centred at 9 h (G09), 12 h (G12) and 14.5 h (G15).

The photometry for the KiDS imaging data has been processed independently by GAMA \citep{KiDSBellstedt2020}. Source detection, segmentation, and photometry is done using \texttt{ProFound} \citep{Robotham2018}. Compared to for example Source Extractor \citep{BertinArnouts1996}, the key features of \texttt{ProFound} include: improved background characterisation, a watershed deblending algorithm, `segment'-based rather than circular/elliptical apertures, and iterative aperture dilation \citep[][]{Robotham2018, KiDSBellstedt2020, Driver2021}. Each of these features is designed to yield robust measures of the total flux in each band, including blended and crowded sources. 

For our purposes, another key feature of the GAMA catalogue is the effort that has gone into visually inspecting and manually correcting the \texttt{ProFound} segmentation maps, and especially larger galaxies that are overly fragmented or shredded. Of 75863 $r < 20.5$ sources that were visually inspected, 6855 required some level of correction \citep{KiDSBellstedt2020, Driver2021}. We note that overly-deblended or shredded galaxies would appear in our analysis as an excess concentration of faint sources in close proximity to low-redshift galaxies, with the potential to artificially inflate the inferred luminosity function at the lowest redshifts and faintest magnitudes. By reducing, if not eliminating, this problem, the close-checked GAMA deblend/segmentation solutions minimises the potential for such a bias.

\subsection{Sample Definition}
\label{sec:selection}
We follow the basic quality control measures necessary for the GAMA photometric catalogues \citep[see][for details]{KiDSBellstedt2020, Driver2021}, including the removal of duplicates and use of the catalogue's \texttt{class} diagnostic to exclude artefacts, including ghosting and reflections. We adopt the GAMA survey footprint, as defined by the combination of the \texttt{mask} and \texttt{starmask} flags. The first of these flags defines the GAMA survey region; the second excludes areas where source detection and/or photometry may be badly affected by bright stars. With these selections the effective survey area is 54.93, 57.44, and 56.93 square degrees for G09, G12, and G15, respectively, and 169.31 square degrees in total \citep[][]{KiDSBellstedt2020}.

GAMA uses a combination of $r$ magnitude versus $(J-Ks)$ colour and $r$ magnitude versus effective size diagnostics to classify detections into categories of \texttt{artefact}, \texttt{star}, \texttt{galaxy}, or \texttt{ambiguous} (where the two star/galaxy diagnostics suggest conflicting classifications). \cite{Driver2021} suggests this \texttt{ambiguous} population is mostly but not entirely made of up stars, and will include unresolved sources with non-stellar colours as well as, e.g., binary stars, and with increasing photometric scatter for the faintest magnitudes. As shown in Appendix \ref{app:stellar_contribution}, our analysis is robust to the presence of stars, quasars, or any other real or artificial source population that do not follow the large-scale structure as traced by the reference sample: our results and conclusions do not change significantly if we include artefacts and stars. We therefore use the GAMA \texttt{class} to exclude artefacts and stars, to limit their potential to slightly increase the statistical errors in our main analysis. We choose to retain \texttt{ambiguous} sources, however, to minimise any selection effects against small/unresolved galaxies.

\subsection{Magnitude and Surface Brightness Selection Limits}
\label{sec:limits}
\begin{figure*}
	\centering
	\includegraphics[width=\textwidth]{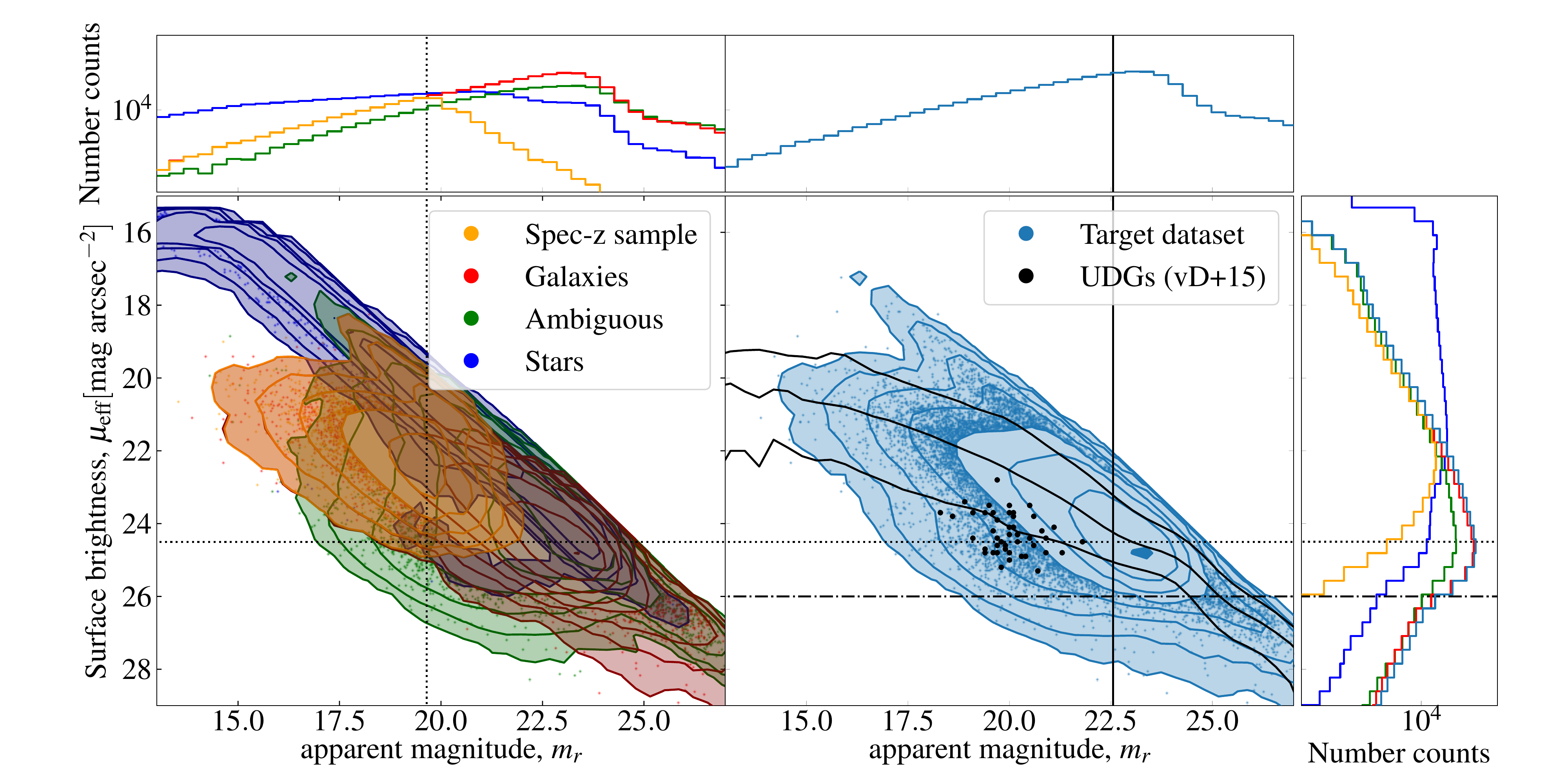}
	\caption{Bivariate brightness distribution (BBD) for the $r$-band parent sample from GAMA photometry of KiDS imaging. In the left panel points are colour-coded by photometric classification: stars (dark blue), galaxies (red), and ambiguous (green) after excluding artefacts. We also highlight the GAMA spectroscopic redshift reference sample in orange. The completeness limit of the spectroscopic sample at $m_r \sim 19.65$ is shown as a vertical dotted line. In the right main panel the BBD for the target dataset is shown (light blue). The three black lines indicate the 5th, 50th and 95th percentile points of the SB distribution depending on magnitude. We chose a $m_r < 22.55$ limit to ensure that incompleteness in our faintest magnitude bins was limited to a few percent for the photometric sample which is shown as a vertical black line. In addition we show the surface brightness limit of the SDSS imaging at $\mu_{eff}=24.5$ corresponding to the spectroscopic sample and the SB limit of our dataset at $\mu_{eff} \sim 26$. In the second panel the positions of 47 low surface brightness objects by \protect\cite{Dokkum2015} (vD+15) are displayed as black points, which are clearly included within the limits of our study, showing that our analysis is sensitive to UDGs.}
 	\label{fig:Surface-Brightness}
\end{figure*}

Fig. \ref{fig:Surface-Brightness} shows the joint $r$-band magnitude-surface brightness distribution, with points colour-coded according to their photometric \texttt{class}. We also highlight the GAMA spectroscopic reference sample. These points are not sharply bounded to the original $m_r < 19.8$ selection limits because their magnitudes have been updated by the improved data and photometry \citep[see][]{KiDSBellstedt2020}. In this diagnostic plot, point sources fall along the linear track as traced by the stars, with the turnover at very bright magnitudes showing the saturation limit in the KiDS imaging. The \texttt{ambiguous} population can be seen to be largely, but not exclusively, extending the stellar population to fainter magnitudes. It is also clear, however, that there is an increasing number of ambiguous sources that coincide with the galaxy population.

Given our focus on the faintest galaxies at low redshift, our analysis will be limited by the depth of the photometric parent catalogues. The faint magnitude limit of the catalogue can be gauged by considering the point where the number counts start to plateau and fall away at $m_r \approx 23$. That this is considerably brighter than the 5$\sigma$ limit for point sources reflects the dominance of extended sources at these faint magnitudes.

It is challenging to meaningfully quantify the limiting surface brightness, which depends on the peak surface brightness averaged over the PSF-scale, modulo details of the \texttt{ProFound} detection algorithm and parameters. What we can see from this diagnostic is that bulk of the our target population is seen with effective surface brightness $\mu_{eff} \lesssim 26$. To gauge where surface brightness selection effects start to significantly bias our otherwise magnitude-limited sample, we have considered how various percentile points of the SB distribution for galaxies vary as a function of magnitude. What we see is that the median and 95th percentiles track roughly linearly down to $m_r \approx 23$, after which point the distribution can be seen to taper towards fainter magnitudes. This also coincides with a mild flattening of the median point of the SB distribution. Both the tapering of the observed distribution and the levelling off of the median are indicators that the low surface brightness tail of the distribution is being missed. We do see some narrowing between the 95th and 99th percentiles over the range $21 \lesssim m_r \lesssim 23$ (not shown), which might be taken to indicate incompleteness at the level of a few percent at these magnitudes.

With these considerations, we limit our analysis to $m_r < 22.55$, as the point where SB selection effects are minimal: not more than a few percent. Beyond this, we make no attempt to correct or account for SB selection effects, noting that any incompleteness will mean that our results are an underestimate of the true population. Our limits show that we are even sensitive to very low surface brightness objects as ultra-diffuse galaxies (UDGs) \citep[see][]{Dokkum2015}. In total $\sim3\times10^6$ target sources are within our study. 

\subsection{Spectroscopic Redshifts for the Reference Sample}
\label{sec:speczs}
The method of clustering redshift inference requires having a reference set with known redshifts/distances to trace the large-scale structure across the target area. We use the GAMA spectroscopic redshift survey for this purpose. GAMA was a multi-year campaign with the 3.9m Anglo Australian Telescope (AAT). At survey end, GAMA had achieved $> 99$\% redshift completeness to the original $m_r < 19.8$ selection limit over each of G09, G12, and G15, with no discernible bias as a function of pair separation \citep{Liske2015}. As discussed above, \citet{KiDSBellstedt2020} and \citet{Driver2021} have described updates to the GAMA photometric catalogues, including re-linking spectra and redshifts to photometric objects. With the updated photometry, which recovers additional flux beyond Source Extractor's AUTO aperture, the 95\% redshift completeness limit dropped to $m_r\sim19.65$.  

As a reference sample for clustering redshift inference, the most pertinent aspects of the GAMA sample are the source density and the redshift interval ($0 < z \lesssim 0.5$ with median redshift $\approx 0.22$). After basic quality control to ensure robust redshift measurements ($nQ > 2$), we have $\sim 170,000$ spec-$z$ measurements. We note that, just as our clustering redshift analysis is robust to `interlopers' in the target sample, the analysis is virtually insensitive to redshift blunders. We also note that, for the purposes of redshift inference itself, it is not necessary or even desirable for the reference sample to be complete or representative. As will be discussed in Sec.\ref{sec:spec-$z$s-like}, we also use the GAMA spectroscopic redshift sample to constrain the overall normalisation of our clustering redshift measurements, via the value of the characteristic density, $\phi_0^*$. Here the completeness of the magnitude-limited GAMA sample is very valuable.

\section{Redshift information from clustering}
\label{sec:cluster-$z$s}
Clustering-based redshift inferences (cluster-$z$s) provide an avenue to statistical redshift information for an ensemble of target objects, based only on positional information. Cluster-$z$s work by cross-correlating the positions of the target sample with the positions of a reference sample for which redshifts are known. By computing the relative strength of the 2D angular cross-correlation for sub-samples of the reference set binned by redshift with the target ensemble, it is possible to infer the target redshift distribution. Unlike spectroscopic redshift measurements (spec-$z$s) or photometric redshift estimates (photo-$z$s), cluster-$z$s do not give redshift information for individual objects, but instead a redshift distribution and source galaxy bias for an ensemble.

Since the only requirement for this method is positional information, it is applicable in regimes where spec-$z$ and photo-z approaches are impractical or even impossible (especially for faint and/or nearby sources, where photo-$z$ errors become comparable to the redshift values themselves). At fainter magnitudes, where conventional object-by-object approaches are more expensive and less reliable, the cluster-$z$s becoming increasingly  useful as the number of sources (and so the statistical significance of the clustering signal) grows rapidly. 

For this method we need three datasets (see Sec. \ref{sec:data}). Firstly the target dataset: it consists of objects for which the cluster-$z$s are calculated solely using their angular positions on the sky ($ra$, $dec$). 

Secondly a reference dataset, mapping the cosmic skeleton, is needed. This set has to consist of objects with accurate measurements of their full 3D position ($ra$, $dec$ and $z$). The objects in the reference set are not required to be of the same type, magnitude, colour, etc. as the target galaxies. The only additional requirement for the reference sample is that it overlaps the region of the target sample. The redshift range of the resulting cluster-$z$s is solely limited by the reference sample and its density. The statistical power is limited by the number of reference objects and the number of targets, which leads to generally better constraints for fainter magnitudes due to larger number counts.

In addition to the two samples mentioned above, an unclustered random sample, which covers the same area and the same angular distribution of the reference sample, has to be generated during the calculation. It is used to measure the auto-correlation function of the reference sample in order to estimate its galaxy bias (see next section for more details). In order to reduce noise, we use more than 100 times as many random data points as reference points.

The overall process of deriving cluster-$z$s is illustrated for three redshift slices for each of the three GAMA regions in Fig. \ref{fig:Cluster-Explanation}. In the left three panels the target data-points  overlay the reference data-points of three different redshift slices. Secondly the corresponding cross- ($w_{tr}$) and auto-correlation ($w_{rr}$) functions are calculated based on the datasets in the first panel. The resulting functions are shown in the middle panels within the separation ranges. By summing $w_{tr}$ and $w_{rr}$ within a certain clustering range the final cluster-$z$ amplitude at each redshift slice is calculated. If this value is calculated for all redshift slices, the final cluster-$z$ distribution can be constructed. The details of this process are explained in the following sub-sections.

\subsection{Cluster-$z$s formalism}
\label{sec:cluster-$z$s-calc}
As described in detail by e.g. \cite{Menard2013}, clustering-based redshift inference works by considering the parameter $\bar{w}_{t}$, called the clustering amplitude. $\bar{w}_{t}$ is obtained as an integral of the angular cross-correlation function of the target and reference samples over a certain angular range $\theta$ limiting the measurement to certain physical scales:
\begin{align}
\bar{w}_{t}(z)=\int_{\theta_{\min }}^{\theta_{\max }} d \theta W(\theta) \, w_{t}(\theta, z)
\label{eq:clusterint}
\end{align}
The usual choice of weight $W(\theta) \propto \theta^{-1}$ maximises the signal-to-noise ratio assuming measurements are Poisson-noise limited. For a fair comparison between sources as a function of redshift, the integration should be done over a fixed projected separation range $r_c$, rather than a fixed angular range. The lower limit of the integration should be large enough to exclude self-correlations of individual sources, avoid fibre collision, ensure that a deterministic galaxy bias model applies \citep[][]{Swanson2008} and not too large that genuine associations are missed. The upper limit should be large enough to capture the LSS, but not so large that statistical noise is added due to uncorrelated background galaxies. The optimal integration limits of $r_{c}$ depend on the particular target and reference samples. As we are probing the correlation of galaxies on small scales we choose $1 \, \mathrm{kpc}/h < r_{c} < 250 \, \mathrm{kpc}/h$. While this choice is important our results are not particularly sensitive to the exact values, and we have tested that none of our results or conclusions change significantly with the integration limits.

As derived in \cite{Menard2013}, the clustering amplitude as defined in Eq.\ \eqref{eq:clusterint} can be related to the underlying clustering bias and redshift distribution as:
\begin{align}
\bar{w}_{tr}(z) \propto \frac{d P}{d z}(z) \bar{b}_{t}(z) \bar{b}_{r}(z) \bar{w}_{m}(z) ~ .
\label{eq:w_tr}
\end{align}
In words: the spatial cross-correlation of the target data with the reference data is the product of: the galaxy bias factors of the reference and the target sample, $\bar{b}_{r}(z)$  and $\bar{b}_{t}(z)$, respectively; the dark matter clustering amplitude $\bar{w}_{m}(z)$; and the shape of the redshift probability distribution $\frac{d P}{d z}(z)$. The bars above each variable represent that they have been integrated within the range $r_c$ as shown in Eq. \ref{eq:clusterint}. Likewise, $\bar{w}_{rr}(r_c,z)$ can be determined from the auto-correlation function of the reference sample. Assuming that the variation of galaxy bias within the clustering range is negligible, $\bar{w}_{rr}(r_c,z)$ can be expressed as:
\begin{align}
    \bar{w}_{rr}(r_c,z)&=b^2_r(z)\bar{w}_m(z)/\Delta z ~ .
    \label{eq:w_rr}
\end{align}

If within the redshift range $\Delta z$ the relative variation of $\frac{d P}{d z}(z)$ dominates over $\bar{b}_t(z)$, we approach the regime where $\frac{d P}{d z}(z) \rightarrow P(z)\delta_D(z-z_0)$ \citep[][]{Menard2013}. Hereby the redshift probability distribution of the target sample can be obtained up to an unknown normalisation which depends in detail on the unknown, and in general evolving, bias of the target sample:
\begin{align}
P_{m,z} \propto \frac{\bar{w}_{tr}}{ \sqrt{\bar{w}_{rr} \Delta z}}\times \frac{1}{\overline{b}_t(z)\sqrt{\bar{w}_m(z)}} ~ .
\label{eq:Pz}
\end{align}

\subsection{Observables}
\label{sec:cluster-$z$s-observables}
The quantities that appear in Eq. \ref{eq:Pz} cannot be measured directly; instead, pair counts are used to estimate the correlation functions. The estimator for the cross-correlation clustering amplitude is given by \cite{Peebles1974},
\begin{align}
\bar{w}_{tr}(r_c,z)=\frac{N_{Rr}}{N_{Dr}}\times\frac{D_tD_r}{D_tR_r}-1
\label{eq:cross}
\end{align}
and the estimator for the auto-correlation of the reference sample by \cite{Landy1993},
\begin{align}
\bar{w}_{rr}(r_c,z)=\frac{N_{Rr}^2}{N_{Dr}^2}\times\frac{D_rD_r}{R_rR_r}-2\times\frac{N_{Rr}}{N_{Dr}}\times\frac{D_rR_r}{R_rR_r}+1
\label{eq:auto}
\end{align}
The notation $XY$ represents the angular cross-pair count across the two datasets $X$ and $Y$, and $XX$ represents the angular auto-pair count within the dataset $X$. The normalisation $N_X$ and $N_Y$ corresponds to the number of points in the datasets $X$ and $Y$. In our case the target dataset is represented as $D_t$, the reference set as $D_r$ and the random data as $R_r$. The normalisation factors are labeled accordingly. For the measurement of the angular pair-counts needed in both estimators we use the python package \texttt{corrfunc} \citep{corrfunc2017}.

In order to get an estimate of the uncertainty in the measurement, errors are calculated via bootstrapping of 30 samples of the reference dataset. For each sample the clustering redshift estimate is calculated in the  same way as the measurement itself. After considering and testing multiple approaches, we applied the normalised median absolute deviation (NMAD) as our final technique to calculate the standard deviation in each redshift bin. We inspected the bootstrapped distributions to check that the assumption of Gaussianity is reasonable.

\subsection{Redshift inference}
The quantities described in the previous sections can be evaluated for the whole sample or any subset. By performing this calculation in redshift bins for different subsets of the reference sample, we are able to build the full redshift distribution $P_z$. If the target sample in addition is split into subsets by magnitude, we can obtain the information needed to measure the GLF.

\begin{figure*}
	\centering
	\includegraphics[width=\textwidth]{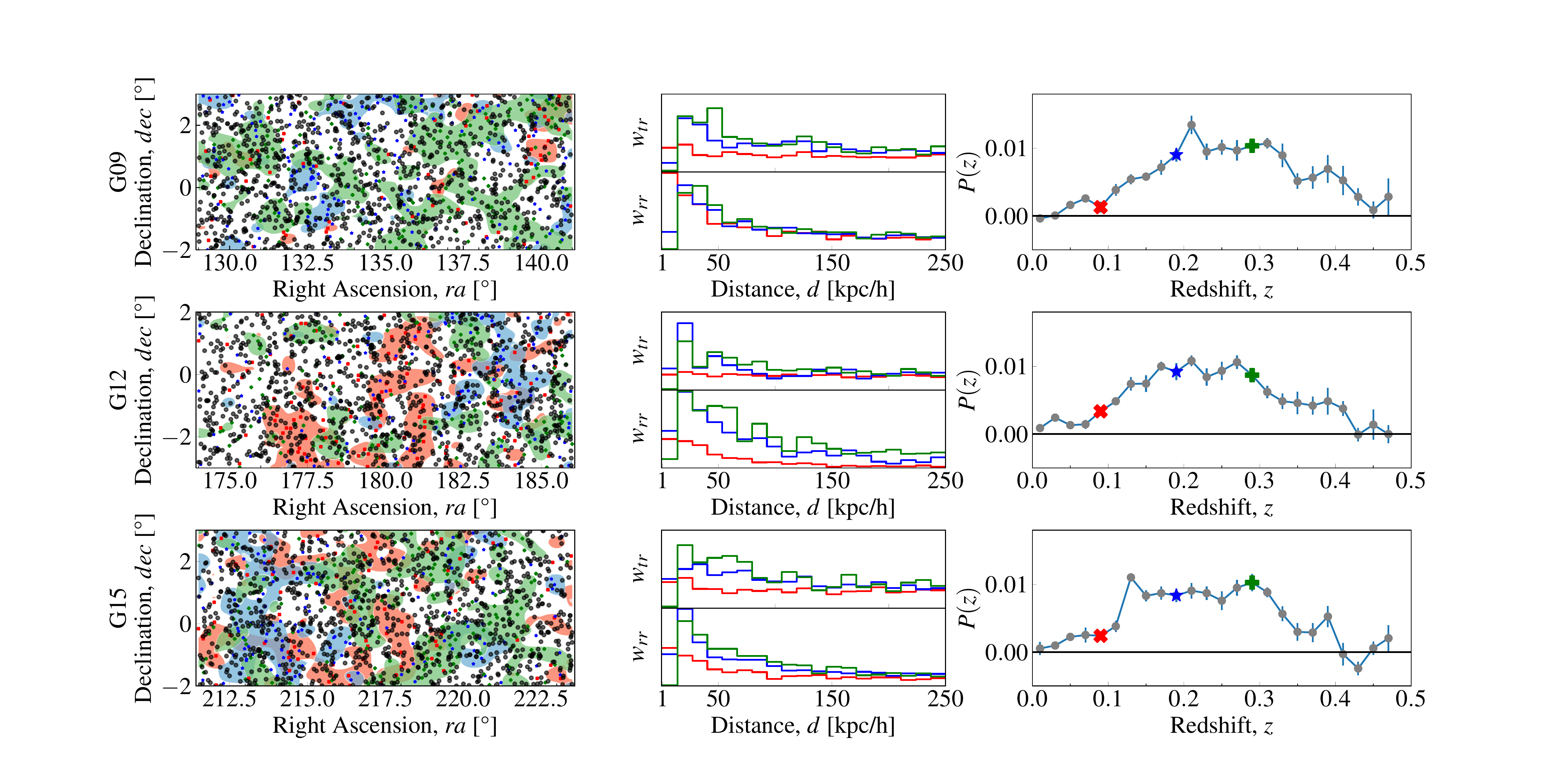}
	\caption{Illustration of the clustering redshifts process. In the left panel the contours of three different redshift slices at $z=\{0.09 \text{ (red)},0.19 \text{ (blue)},0.29 \text{ (green)}\}$ with $\Delta z = 0.02$ is shown, on top of a sub-sample of the target data. Here the differences of the cosmic web at these redshifts can be clearly seen. These differences are used to calculate the cross-correlation of the target dataset and the reference data at these redshifts. The resulting cross-correlation function $w_{tr}$ and the auto-correlation function $w_{rr}$ are shown in the diagrams in the middle. By using the summed results of these functions over the corresponding clustering ranges $r_c$, the $P_{m,z}$ at these redshifts are calculated. The resulting $P_{m,z}$ is shown in the last panel and the points derived at the three redshift slices are marked accordingly.}
	\label{fig:Cluster-Explanation}
\end{figure*}

The magnitude and redshift bins should be as small as possible for the best resolution, but large enough to ensure a statistically significant measurement for each redshift/magnitude cell. For our sample, $\Delta z = 0.02$ and $\Delta m = 0.25$ is a good compromise between these two considerations.

\subsection{The need for the normalisation factor $A_m$}
\label{sec:Afac}
The output of the cluster-$z$s measurement is only proportional to the redshift distribution, $N(z)$. A normalisation is therefore needed to compare the output to the data, or to derive the GLF. The difficulty at this point is that a simple normalisation by the number of target data points is not reliable because, first, it is unknown if all target objects are truly galaxies, and any additional contribution from, e.g. stars or quasars, would result in an overestimation of the final $N(z)$. The second aspect is that a certain fraction of objects in the target dataset might reside beyond the limits of the reference set. This would again result in an overestimation of the $N(z)$. Therefore an important  challenge  is how to tackle the derivation of the normalisation factors, despite these uncertainties.

Mathematically speaking, the output of the cluster-$z$s process $P_{m,z}$ is proportional to the true number counts $N_{m,z}$. A normalisation parameter $A_m = P_{m,z} / N_{m,z} $ is therefore necessary to transform the cluster-$z$s into proper number counts $N_{m,z}$, and vice versa. Given an expectation of the true $N_{m,z}$ we are able to find the best-fitting value of $A_m$ that is consistent with both $N_{m,z}$ and our clustering measurements $P_{m,z}$. In this situation, the least squares or maximum likelihood estimate for $A_m$ is analytic:
\begin{align}
\chi^2 &=  \sum \frac{(A \times \Phi(z) - P_z)^2}{\sigma_{P_z}^2}  \\
\frac{\partial \chi^2}{\partial A} &\overset{!}{=}0 \\
\Rightarrow A &=  \frac{\sum_z \Phi(z) \times P_z / \sigma_{P_z}^2 }{\sum_z \Phi(z) \times \Phi(z) / \sigma_{P_z}^2 }
\label{eq:A-derivation}
\end{align}
$A_m$ is therefore the maximum-likelihood solution for the normalisation given the model and the data, where $\Phi$ is the expected $N_{m,z}$ from a model calculated for a certain set of parameters or binned spec-$z$s data, $P_z$ is the unnormalised data, e.g. the cluster-$z$s, and $\sigma$ is the corresponding standard deviation of the data.

\subsection{Bias evolution}
\label{sec:bias}
The main limitation of the cluster-$z$ approach is the unknown bias evolution of the target sample, which is degenerate with the inferred redshift distribution. A constant factor, assuming an evolution of the target sample galaxy bias over redshift in a way which cancels out the growth of the dark matter structure, is of no concern as it can be absorbed into the normalisation scalar. A larger concern would be variations in the mean target bias over redshift.

Eq. \ref{eq:Pz} shows how the effect of a varying $\bar{b}_t(z)$ changes the shape of the inferred redshift distribution: i.e.,\ $P(z) \propto \bar{b}_t(z)^{-1}$. One way to mitigate this issue is to preselect target samples over narrow redshift intervals (eg. using photo-zs) to minimise any differential bias across each sample. We have chosen to assume a model where the growth factor in combination with the unknown bias is constant $\bar{b}_t(z)\sqrt{\bar{w}_m(z)} = const$. This choice is based onto the assumption that the unknown bias $\bar{b}_t(z)$ is increasing with redshift, whereas $\sqrt{\bar{w}_m(z)}$ decreases with redshift. The unknown bias evolution remains our main systematic error and limitation. To mitigate its impact on our results, we focus on the low-z GLF.

\subsection{Validation of the cluster-$z$s}
\label{sec:cluster-$z$s-validation}
In order to validate the cluster-$z$s process, we tested our ability to recover the known redshift distribution in bins of apparent magnitude of the GAMA sample for each region. 
While this test uses the same dataset for both the target and reference samples, we stress that this test is not circular. Firstly, for the calculation of the $P_{m,z}$, all points within a magnitude bin (target objects) are correlated to all points within a redshift bin (reference data).

\begin{figure}
    \centering
    \includegraphics[width=\columnwidth]{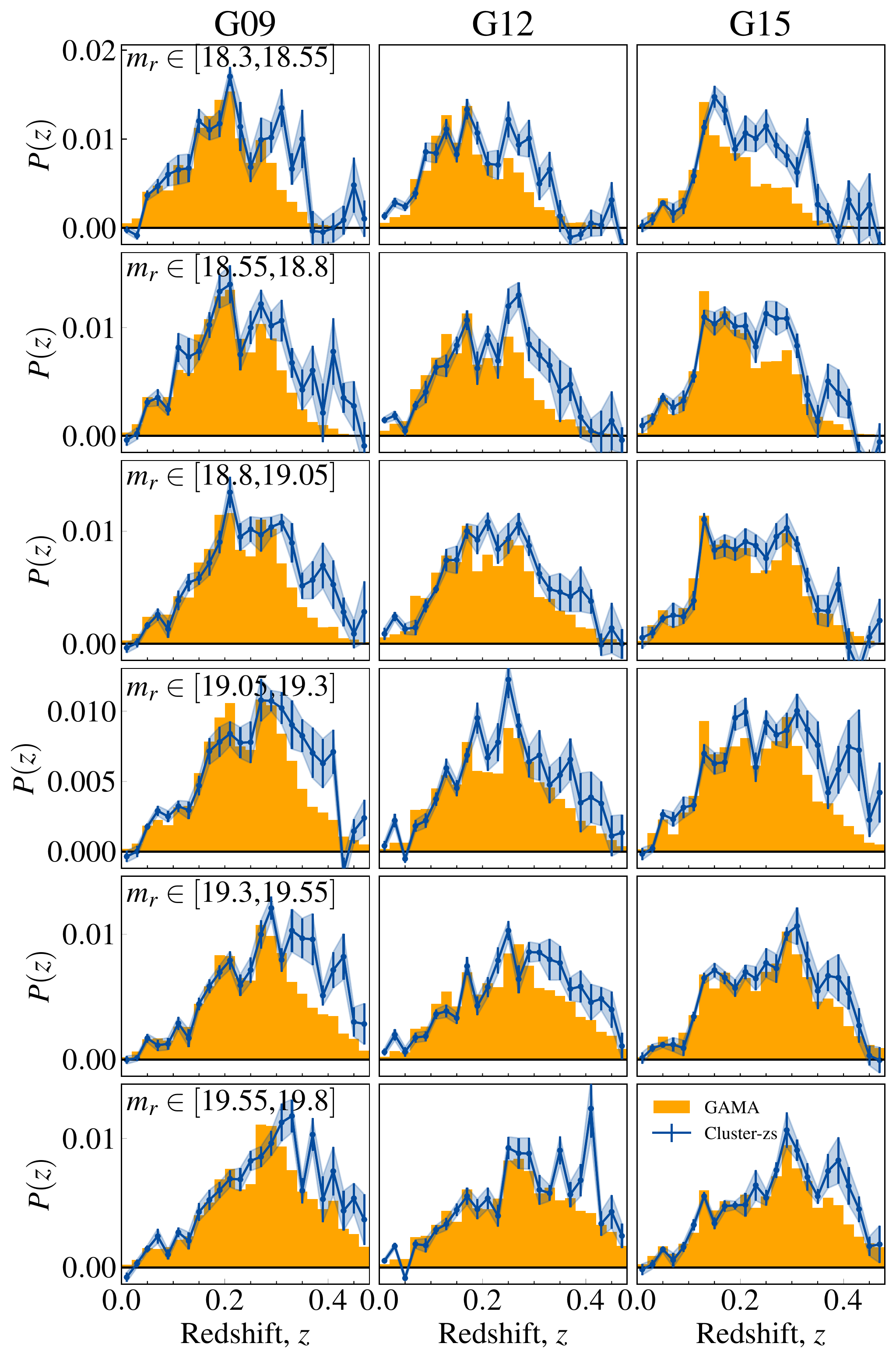}
    \caption{Comparison of the derived cluster-$z$s for each GAMA region in blue to the spec-$z$ distribution of the corresponding GAMA galaxies in orange, normalised by the $A_m$s for different magnitude bins. It can be seen that the cluster-$z$s recover the spec-$z$ distribution at low-z to intermediate-z, and the discrepancy increases with redshift due to the evolving galaxy bias.}
    \label{fig:clusterzs}
\end{figure}

In Fig. \ref{fig:clusterzs} it can be seen that the resulting redshift probability distribution is in general in good agreement with the GAMA spec-$z$s for redshifts $z < 0.3$. The cluster-$z$s follow the spec-$z$ distribution and even reproduce some of the large-scale structure features unique to each GAMA region. Given the fluctuations between each region, we use all three regions for our study in order to reduce the errors due to field-to-field variance.

At redshifts $z > 0.3$ the cluster-$z$s tend to overestimate the spec-$z$ distribution. It is conceivable that this reflects some incompleteness in the spectroscopic redshift catalogues that preferentially acts against higher redshift galaxies; e.g.\ surface brightness effects, blending/confusion in the input catalogues, etc. The other alternative is that this reflects the evolving differential bias between the redshift-binned reference sample and the magnitude-binned target sample. One way to test this is by correcting the resulting cluster-$z$s using the auto-correlation function for the target data similar to the correction of the reference bias, which does improve the correspondence between spec-$z$ and cluster-$z$ results. We can also fit for a bias correction of the form $\bar{b}_t(z) = (1+z)^\beta$ by requiring that the spec-$z$ and cluster-$z$ distributions agree: we find $\beta \approx 1$. This shows the impact of differential bias as the dominant source of systematic error, especially for $z \gtrsim 0.3$, but unfortunately all these corrections are only possible for the brighter magnitude bins, where spec-$z$s are available, but as our main focus is the low-z GLF we remain with our constant assumption. We return to this issue in Sec. \ref{sec:DoesMatter}.

The errors in the cluster-$z$ measurement, derived from bootstrapping, can be seen to increase with redshift. This behaviour follows the redshift distribution of the reference dataset and the larger uncertainties are a result of the decreasing number counts at higher redshift.  In addition it can be seen that at brighter magnitudes the cluster-$z$s can produce negative values at high redshifts. These negative correlation amplitudes, originating from statistical noise and systematic effects, highlight the point that the output of the clustering methodology is only similar to a normalised probability function, but not the same.

\section{Modelling and Measuring the GLF with cluster-$z$s}
\label{sec:fitting}
The goal of this paper is to use the clustering redshift measurements described in Sec. \ref{sec:cluster-$z$s} to determine the luminosity function for $z \sim 0$ galaxies to the faintest possible limits. In principle, the evolving luminosity function $\Phi(m|z)$ can be directly inferred from the observed bivariate distribution $N_{m,z}$ plus cosmology. The main obstacle is that clustering redshift inference yields only the shape of the redshift distribution: our clustering redshift results are only proportional to the true redshift distribution up to an unknown normalising scalar; i.e. $P_{m,z} = A_m \times N_{m,z}$. In this section we describe our process for determining these normalisation factors, $A_m$, that relate our cluster-$z$ measurements to the underlying redshift distribution, and hence to the true galaxy luminosity function.

Our solution is to use a parametric model for the evolving luminosity function, as described in Sec. \ref{sec:simple_model}, to describe the shape of the redshift distributions, $N_{m,z}$. The parameters of this model are constrained by our cluster-$z$ results, as described in Sec. \ref{sec:spec-$z$s-like}. For any choice or trial set of model parameters, the maximum likelihood estimate of the factors $A_m$ is analytic. The one complication is that we need an external constraint on the global normalisation of the model, which is otherwise degenerate with the $A_m$s. Sec. \ref{sec:spec-$z$s-like} describes how we use the GAMA spectroscopic sample to break this degeneracy. Since the values for $A_m$ can be computed for any specified choice of parameter values, the same likelihood analysis can be used to give the posterior probability distribution functions (PDF) for the model parameters and also for the set of scalar $A_m$s.

\subsection{A simple model for the evolving GLF}
\label{sec:simple_model}
The characteristics of the observed GLF are a power law slope for fainter magnitudes with an exponential drop-off at bright magnitudes, which are usually described using a \cite{Schechter1976} function:
\begin{align}
S(M|M^*,\alpha,\phi^*) =  & 0.4 \ln 10 \phi^{*}\left[10^{0.4\left(M^{*}-M\right)}\right]^{\alpha+1} \nonumber\\
&\times \exp \left[-10^{0.4\left(M^{*}-M\right)}\right] \mathrm{d} M
\end{align}
where $\phi^{*}$ is the characteristic number density, $M^*$ the characteristic magnitude cut-off and $\alpha$ defines the faint-end slope of the function. Recent studies have shown that a single Schechter function does not provide a good description of the GLF across a broad range of magnitudes on its own, and instead the use the sum of two Schecter functions \citep[e.g.][]{Baldry2008, Moffett2016} is necessary. There are theoretical arguments to understand this double Schechter form for the GLF in terms of the relative efficiency of mass- and environment-dependent feedback processes \citep[e.g.][]{White1978, Kauffmann1993, Cole1994, Peng2010b}.

For the redshift evolution we use the parametrisation by \cite{Lin1999} \citep[see also][]{Loveday2012, Loveday2015} for describing a linear evolution of the logarithmic galaxy density $\log(\phi_i^*)$ and characteristic magnitude $M^*$ using the parameters $Q$ and $P$. The slope $\alpha$ is kept constant in this parametrization.
\begin{align}
M^*(z)&=M^*-Q \times z \\
\phi_i^*(z)&=\phi^*_0 \times 10^{0.4 \times P \times z}\\
\alpha(z)&=\alpha
\end{align}
This simple parametrisation relies on the assumption that the shape of the luminosity function is not evolving and the function is only shifted horizontally in absolute magnitude by $Q$ and vertically in density by $P$. As we are not explicitly considering k-corrections, we rely on $Q$ to absorb their effects and our results should always be understood in terms of the observer-frame r-band. Given our focus on the low-z GLF, k-corrections are of minor importance in our study. The final double Schechter function $S_d$ is therefore given by:
\begin{align}
\Phi(M;M_i^*,\alpha_i,&\phi_i^*,Q_i,P_i) \nonumber \\ = S_1(&M;M_1^*,\alpha_1,\phi_1^*,Q_1,P_1) \nonumber \\ 
& + S_2(M;M_2^*,\alpha_2,\phi_2^*,Q_2,P_2)
\end{align}
We distinguish between the two Schechter functions by requiring $\alpha_1 < \alpha_2$. The model describes $\Phi(M, z)$, while our observation is $N(m,z)$. The two are related via $N(m, z) = \Phi(M, z) \, dV$, with $M = m + DM$, and where cosmology enters via the comoving volume element, $dV$, and the distance modulus, $DM$. In Appendix \ref{app:choice of model} we show that the double Schechter form provides a good description of our data, and also that our main results and conclusions are not particularly sensitive to this choice of parameterisation.

\subsection{Cluster-$z$s likelihood function}
The model provides a prediction for $N(m,z)$ integrated within the grid cell based on a particular choice or trial set of parameter values. This predicted $N(m,z)$ should match our cluster-$z$ measurements up to an unknown normalisation. For a particular model, the MLE for $A_m$ is computed using Eq. \ref{eq:A-derivation}, and the log-likelihood $\mathrm{ln}(\Like_{i})$ associated with this parameter combination follows. Assuming Poisson statistics, so that the statistical uncertainties are normally distributed, and using the $A_m$s as described, the logarithmic likelihood function for the data $P_{m,z}$ given the model $\Phi$ for each individual GAMA region $i$ yields:
\begin{align}
\mathrm{ln}(\Like_{i}) = -\frac{1}{2} \sum_{m,z} \frac{( P_{m,z} - A_m\times \Phi(M^\dagger(z),\phi^\dagger(z),\alpha)_{m,z})^2}{\sigma_{m,z}^2 }  \end{align}
The errors are derived from bootstrap resampling of 30 realisations of the data as described in Sec. \ref{sec:cluster-$z$s-observables}. In principle it is possible to use the MCMC sampler to sample the probability distribution of the $A_m$s as nuisance parameters, but this is unnecessary as the calculation of the $A_m$s from any specific model is analytic and it is therefore faster and easier to parcel this out of the MCMC process.

By modeling the regions simultaneously with their field-specific $A_m$, we account for field-to-field variations. The overall likelihood is then obtained from the product of the individual likelihoods.
\begin{align}
    \mathrm{ln}(\Like_1) =  \ln(\Like_{\text{G09}}) + \ln(\Like_{\text{G12}}) + \ln(\Like_{\text{G15}})
\end{align}
From Eq. 9 and 15 it can be seen that there is a degeneracy between the values of the normalisation factor $A_m$ and the characteristic densities $\phi_0^*$. While this does not impact our ability to constrain the GLF shape using cluster-$z$s, some outside information is therefore required to constrain the overall $z \sim 0$ normalisation of the GLF.

\subsection{The need for a spectroscopic sample to constrain $\phi_0^*$}
\label{sec:spec-$z$s-like}
We use the GAMA spec-$z$ sample to constrain the overall normalisation of the GLF, and so break the degeneracies mentioned above. The simple idea is that the model should explain both the spec-$z$ and the cluster-$z$ results.

While the cluster-$z$ results can only be derived in bins, the spec-$z$ data counts discrete objects. We could bin the spec-$z$ data, but that is not necessary and throws away information. The appropriately normalised model describes the likelihood of observing a data-point at any given point in $(m,z)$ space, thus we are able to evaluate the point-wise likelihood function.

For the fitting of the spec-$z$s we use a point-based likelihood function as described by \cite{Marshall1983}. In this approach the magnitude and redshift plane is split into tiny cells of $dM$,$dz$ which can only contain one or zero objects. The mean number of objects expected in one cell is
\begin{align}
\lambda = \Phi(M,z)\frac{dV(z)}{dz}dz dM \ S(M,z)
\end{align}
where $\Phi$ is the double Schechter luminosity function and $S(M,z)$ is the selection probability, which yields one if an object could be found given the selection boundaries and zero if not. The overall probability, given all galaxies are independent, is the product of the possibilities of having one or zero objects in a bin. Using Poisson statistics this leads to:
\begin{align}
\Like_2=[\prod_i \phi(M_I,z_i)\frac{dV(z_i)}{dz}dzdM][\prod e^{-\phi(M_I,z_i)\frac{dV(z_i)}{dz}S_i dzdM}]
\end{align}
and therefore the log-likelihood of the spec-$z$s becomes
\begin{align}
\ln(\Like_2) = \sum_i \ln[\phi(M_i,z_i)]-\int\int dz dM \phi(M,z) \frac{dV(z)}{dz}
\end{align}
where the second term enforces the integral constraint on the likelihood function, such that a data point must be observed somewhere within the observational window. In this approach we neglect the sample variance contribution to the likelihood function and therefore our errors do not represent field-to-field variations.

\subsection{Using MCMC to condition the model and infer $A_m$}
\label{sec:likelihood}
In the previous three sections we have defined our model for the evolving GLF and the $\mathrm{ln}(\Like)$ function. We now use the MCMC utility \texttt{emcee} \citep{emcee2019}, with uniform priors ($\ln(\Like) = \ln(\Like_1) + \ln(\Like_2) + prior$) on our model parameters, to sample the parameter space subject to the observational constraints. The chains themselves represent the joint PDF for the parameters that define our model for the GLF. To check the convergence of the sampler, the integrated autocorrelation time $\tau$ is calculated as described by \cite{Goodman2010} and the fit is resumed until the estimated autocorrelation time is less than $\tau = N_{samples}/50$.

We recall that each evaluation of the model involves a ML solution for the values of $A_m$, which we record at every step of the chain. These chains represent the joint PDF, incorporating and marginalising over all possible models which are consistent with the data.

\subsection{Obtaining the GLF measurements}
To obtain the GLF measurements, two final steps have to be undertaken after the fit: firstly the cluster-$z$s have to be normalised using the $A_m$s and the best-fitting model. Secondly, we weight the resulting number counts by the cosmological volumes of the corresponding redshift bin and apply the distance modulus to derive the GLF measurements.

Our analysis can thus be understood from two complementary angles. One interpretation would be to emphasise the parametric fits as `the' description of the evolving luminosity function. From this perspective, the set of $A_m$s can be viewed as nuisance parameters, which are a necessary part of the model-to-data comparison, to be marginalised away. Alternatively, a more data-minded approach would view the model as a means of deriving a self-consistent set of values for the critical normalisation factors, $A_m$, from which both $N_{m,z}$ and $\Phi(m|z)$ follow. In this way of thinking, the particular parameter values for the model are less important: what matters most is simply whether the model provides a good description of the underlying data. We defer further discussion of this issue to Sec. \ref{sec:Discussion}.

\section{Results}
\label{sec:results}
\begin{figure*}
	\centering
	\includegraphics[width=0.9\textwidth]{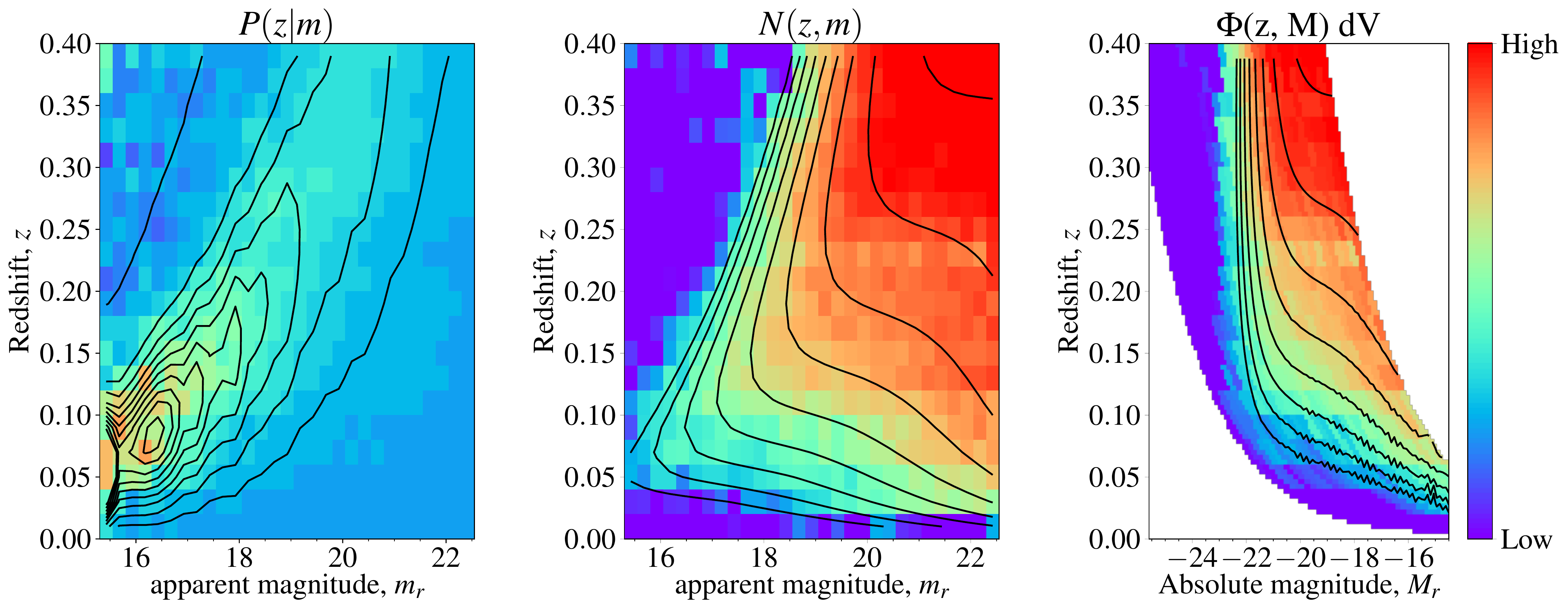}
	\caption{Illustration of the impact of the A-factors. The three panels represent different steps of the analysis. From left to right: a) the unnormalised cluster-$z$s of all three regions, b) cluster-$z$s normalised by the A-factors of the best fit with model contours, c) adding the distance modulus to the A-factor normalised cluster-$z$s with model contours. The colours and contours indicate different levels of a) clustering amplitudes, b) number counts and c) densities with a logarithmic scaling. Moving from a) to b) it can be seen how the normalisation of the cluster-$z$s by the A-factors changes the $P_{m,z}$ of the unknown dataset into a proper $N_{z,m}$. By adding the distance modulus, the absolute magnitude range covered by this study is displayed in c).
	\label{fig:AfacDist} }
\end{figure*}

The full process proceeding from the cluster-$z$s in bins of magnitude to measurements of the GLF is illustrated in Fig. \ref{fig:AfacDist}. In the first panel, the combined raw cluster-$z$s measurements $P(z|m)$ of all three regions are shown. At fixed magnitude, the cluster-$z$ measurements are approximately integral normalised to unity. At bright magnitudes, the relatively narrow distribution of redshifts shows as a relatively strong peak; at fainter magnitudes, the broader redshift distribution is seen as more diffuse in this visualisation. The progression of peak of the distribution shows how the mode of the redshift distribution shifts from bright to faint magnitudes. Besides this tail in the $(m,z)$ plane, there are two regions of interest. Firstly, the amplitudes at higher redshift and bright magnitudes appear noisy. At these magnitudes the number of target sample data-points is small, which lead this area to be noise-dominated. In contrast to that, the region at fainter magnitudes and low z appears almost flat, as at these redshifts the clustering amplitude is small due to accurate clustering-based redshift estimates, and only small numbers of galaxies with the corresponding magnitude are residing at low redshifts.

The cluster-$z$s are transformed into an $N(z)$ using the $A_m$s. In the second panel of Fig. \ref{fig:AfacDist}, the derivation of the $A_m$s from the best-fitting model and its application to the cluster-$z$s can be seen. The contour lines of the different galaxy populations extend with increasing redshift, as more and more volume is covered. In the third panel, and as the last step towards obtaining the GLF, the distance modulus is added to the galaxy distribution. Here the full extent of our study can be seen clearly. We observe how the number of galaxies at higher redshift increases, which is already an indication that there is no flattening of the GLF at the faint end. In the third panel, the jagged model lines result from the pixelisation of the rectangular grid.

The final step of weighting the resulting distribution by the cosmological volumes, and obtaining the GLF, is described in the next section.

\subsection{Parametric description of the evolving GLF}
\begin{figure*}
	\centering
	\includegraphics[width=\textwidth]{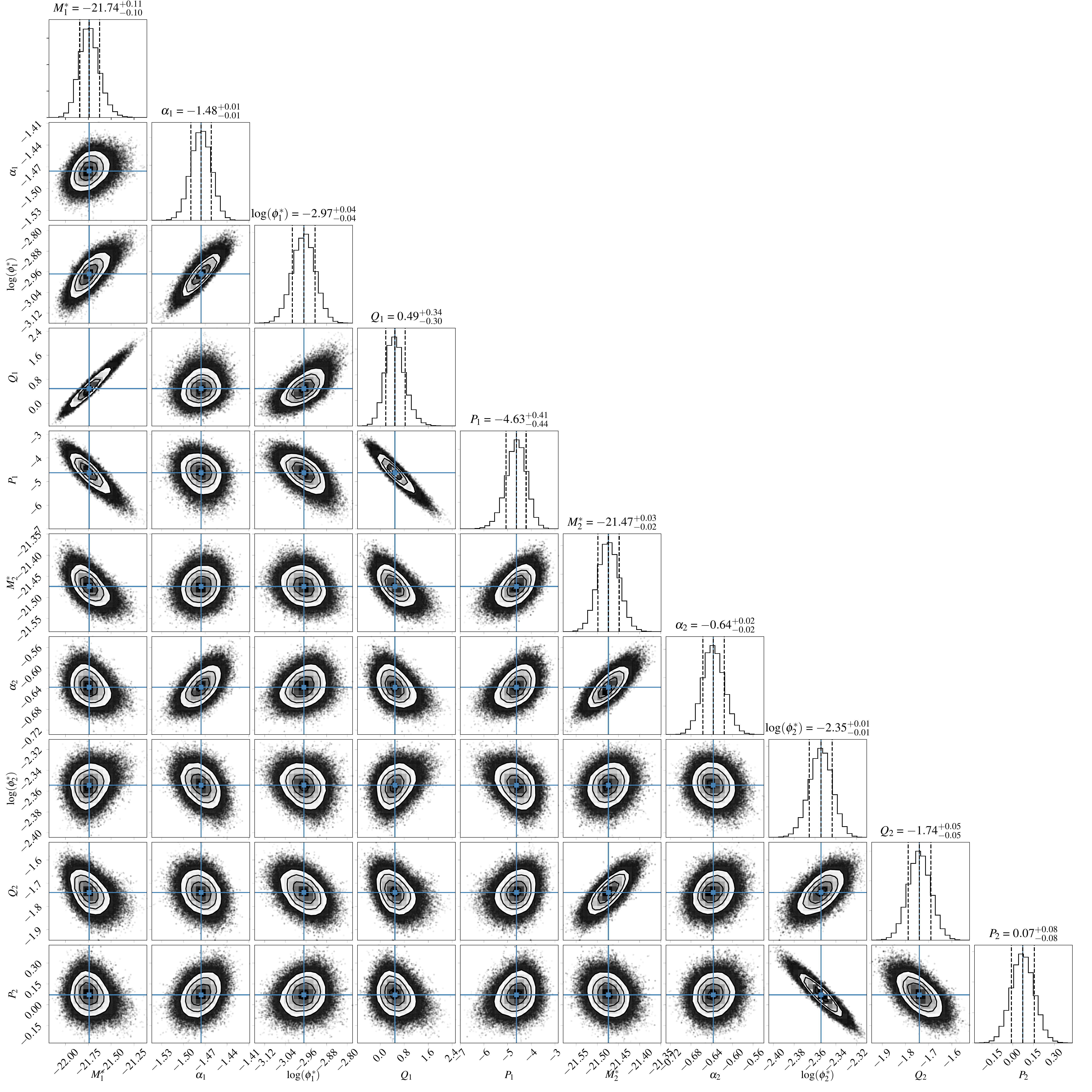}
	\caption{Posterior probability distribution (with uniform priors) for the ten parameters of the double Schechter function fit to the combined cluster-$z$s and spec-$z$ likelihood in the three GAMA fields. Apart from the obvious covariances between the connected parameters, the independence of the two Schechter functions can be seen.
	\label{fig:Corner}}
\end{figure*}

The resulting posterior probability distributions of the fitted model parameters using the KiDS data within $m_r<22.55$ is shown in Fig. \ref{fig:Corner}. It shows the marginalised and joint constraints on the  parameters from our MCMC chains. As defined by the model parameterisation, there are partial degeneracies between the parameters $M^*$ and $Q$, as well as between $\log{\phi^*}$ and $P$. This is evident for both Schechter functions. The resulting probability distributions display the same shape and covariance for both Schechter functions. In addition it can be seen in the central 5x5 cells of the plot that the two Schechter functions are not significantly correlated. The resulting best-fit parameters are displayed in Table \ref{tab:fitparam}, along with the uncertainties derived from the sampler. 

We note that it is common in the literature to fit models including a coupled $M^*$ \citep[e.g.][]{Baldry2012, Wright2017} rather than a decoupled $M^*$ \citep[e.g.][]{Kelvin2014}. We decided to adopt a decoupled $M^*$ as it is more general. With our results we find that the $M^*$s are similar, but not equal. In addition we find that $|\alpha_1 - \alpha_2| = 0.84\pm0.03$ which is close to $\Delta \alpha \sim 1$, which previous studies have measured between early and late type galaxies \citep[e.g.][]{Loveday2012}. In addition the empirical mass-quenching approach by \cite{Peng2010b} produces a Schechter function with common $M^*$ for early and late type galaxies as well as an $\Delta \alpha \sim 1$. 
Here it is perhaps significant that the first component evolves more strongly in magnitude ($P_1 = -4.6 \pm 0.4$) than density ($Q_1 = 0.5 \pm 0.3$), suggesting continued star formation/assembly. In contrast, the second component is growing in density ($Q_2 = -1.74\pm0.05$) but not magnitude ($P_2 = 0.07 \pm 0.08$) suggesting a increasing number of only passively evolving galaxies.
Based on these considerations, it is perhaps tempting to identify our first and second Schechter components as pertaining to the blue/star forming and red/quiescent populations, respectively, even though we have not used any colour or stellar population information in this analysis.

\begin{table}
\centering
\caption{Best-fit parameters of the double Schechter function fit.}
\label{tab:fitparam}
\resizebox{\columnwidth}{!}{\begin{tabular}{lccccc} \topline
 & $M^*$ & $\alpha$ & $\log_{10}(\phi^*)$ & $Q$ & $P$ \dblline
$S_1$ &$-21.74\pm^{0.11}_{0.10}$ & $-1.48\pm^{0.01}_{0.01}$& $-2.97\pm^{0.04}_{0.04}$ & $0.49\pm^{0.34}_{0.30}$& $-4.63\pm^{0.41}_{0.44}$ \midline
$S_2$ &$-21.47\pm^{0.03}_{0.02}$ & $-0.64\pm^{0.02}_{0.02}$& $-2.35\pm^{0.01}_{0.01}$ & $-1.74\pm^{0.05}_{0.05}$& $0.07\pm^{0.08}_{0.08}$ \botline
\end{tabular}}
\end{table}

\begin{figure}
	\centering
	\includegraphics[width=\columnwidth]{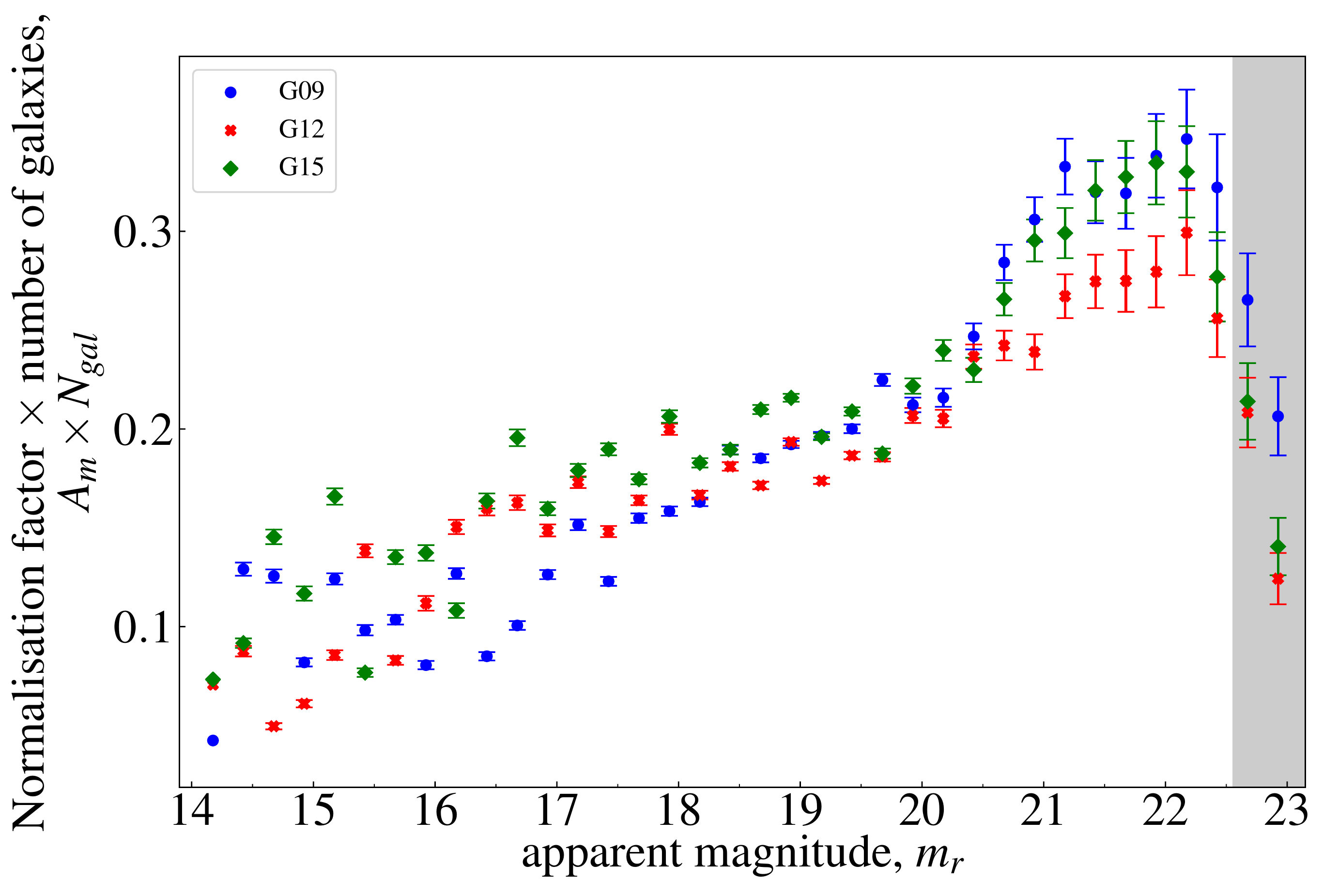}
	\caption{The normalisation factor $A_m$ of each region for each magnitude bin, multiplied by the number of objects in the magnitude bin. The grey area indicates magnitde bins which are potentially incomplete and not part of the main study. It can be seen how the $A_m$s account for the normalisation in the different regimes.}
	\label{fig:Afac}
\end{figure}

Each point in Fig. \ref{fig:Afac} represents a different realisation of the model and shows its corresponding $A_m$. It therefore displays the allowed variation in $A_m$ that is consistent with good but imperfect knowledge of the evolving LF. The $A_m$s incorporate the normalisation due to the increasing number of galaxies (LSS) as well as the magnitude-dependent bias evolution. In order to visualise the effect of the $A_m$s without the different number of objects in each magnitude bin, we display the $A_m$s multiplied by the number of galaxies for each magnitude bin in Fig. \ref{fig:Afac}, which scales as the mean bias multiplied by the variations in LSS for each region in each magnitude bin. The scatter between the points represents the field-to-field variations. At the bright end, where completeness is high and the number of galaxies is low, field-to-field variations are strong. With fainter magnitudes, field-to-field variations become less important. Here it can be seen that the $A_m$s of all three regions follow a linear relationship. The errors of each measurement are underestimated as the sample variance error contribution is neglected.

\subsection{Recovery of the number distribution}
\label{sec:recovery-number-dist}
\begin{figure*}
	\centering
	\includegraphics[width=\textwidth]{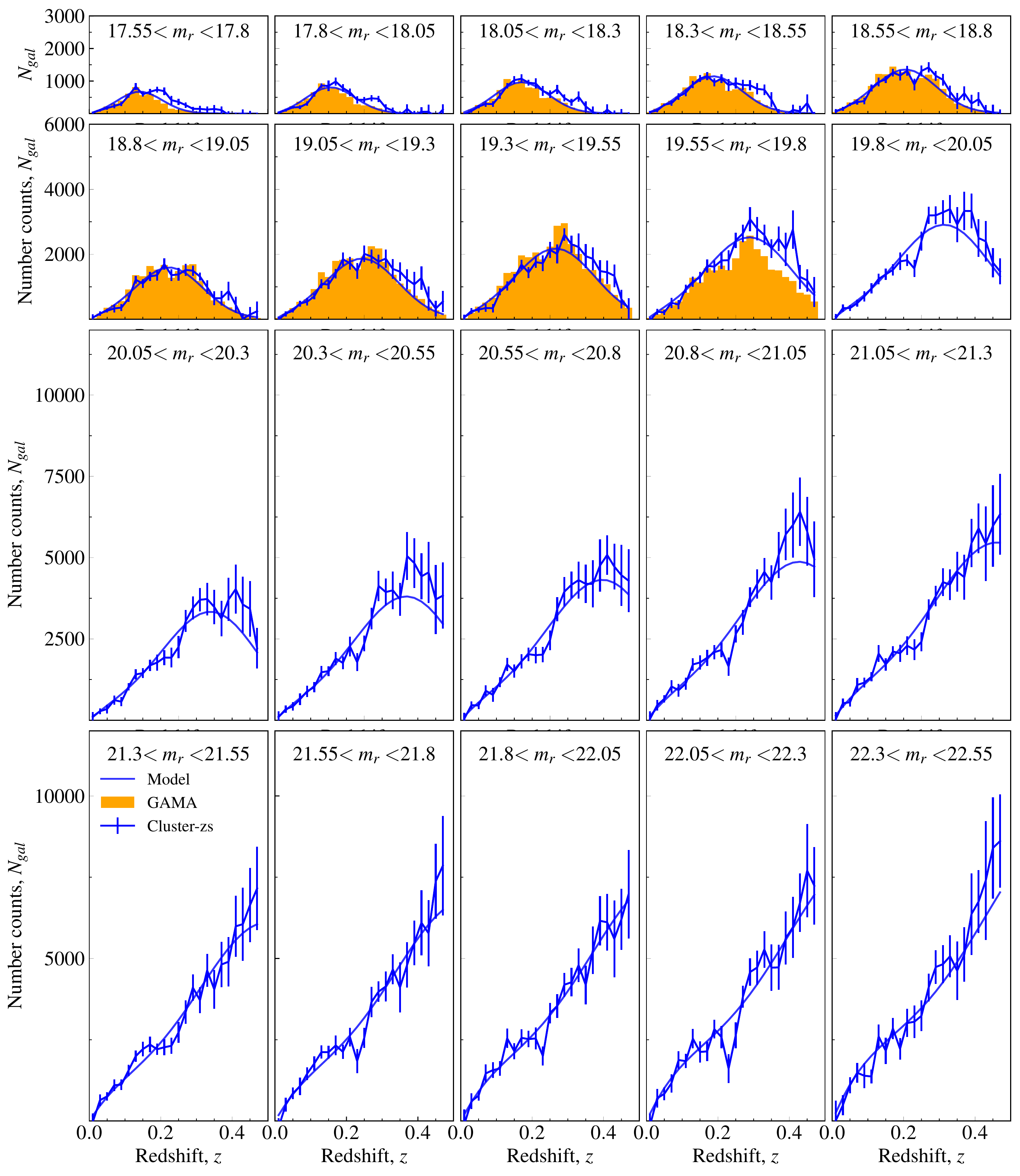}
	\caption{The resulting redshift distributions $N_{z,m}$ from normalising the cluster-$z$s by the $A_m$s (blue lines with error bars) for all magnitude bins is compared to the GAMA spec-$z$ distribution (orange bars) as well as the best-fit model (solid line and shaded errors). The $N_{z,m}$ can be seen to increase with fainter magnitude and to follow the GAMA spec-$z$ distribution, where available. At fainter magnitudes, only a part of the redshift distribution can be traced due to the limitations of the reference set.}
	\label{fig:Nz}
\end{figure*}

Having explored the effect of the $A_m$s, we can now examine the resulting redshift distributions $N_{z,m}$ and compare the normalised cluster-$z$s with the GAMA spec-$z$ distribution. In Fig. \ref{fig:Nz} the resulting $N_{z,m}$ is shown in separate magnitude bins. As the number of GAMA spec-$z$s are significantly dwindling at magnitudes larger than their completeness limit $m_r=19.65$, they are not shown in the diagram for these faint magnitudes. By comparing the GAMA spec-$z$s, shown as bars, with the continuous model lines as well the cluster-$z$s (error bars), a few results can be noted.

Firstly the model with its best fit parameters from Tab. \ref{tab:fitparam} is in good agreement with the spec-$z$s. This model is the basis for the normalisation of the cluster-$z$s. The normalised cluster-$z$s are in general a good approximation for the GAMA spec-$z$s where available. The cluster-$z$s themselves overestimate the true distribution at higher redshifts, as has already been seen in Fig. \ref{fig:clusterzs}, which is due to the unknown target galaxy bias evolution discussed in Sec. \ref{sec:cluster-$z$s}. In contrast to the error in the model, which is rather small especially at brighter magnitudes, the scatter of the cluster-$z$s is always larger, which is emphasising that we are limited by the errors in the cluster-$z$s and not by the scatter in the model or the $A_m$s.

Spectroscopic redshifts dominate at bright magnitudes, at fainter magnitudes objects with redshifts have cluster-$z$s, but the number counts are dominated increasingly by objects beyond our redshift range. For these magnitudes only a part of their redshift distribution can be traced, due to the unavailability of reference points at higher redshifts.

\begin{figure*}
	\centering
	\includegraphics[width=2\columnwidth]{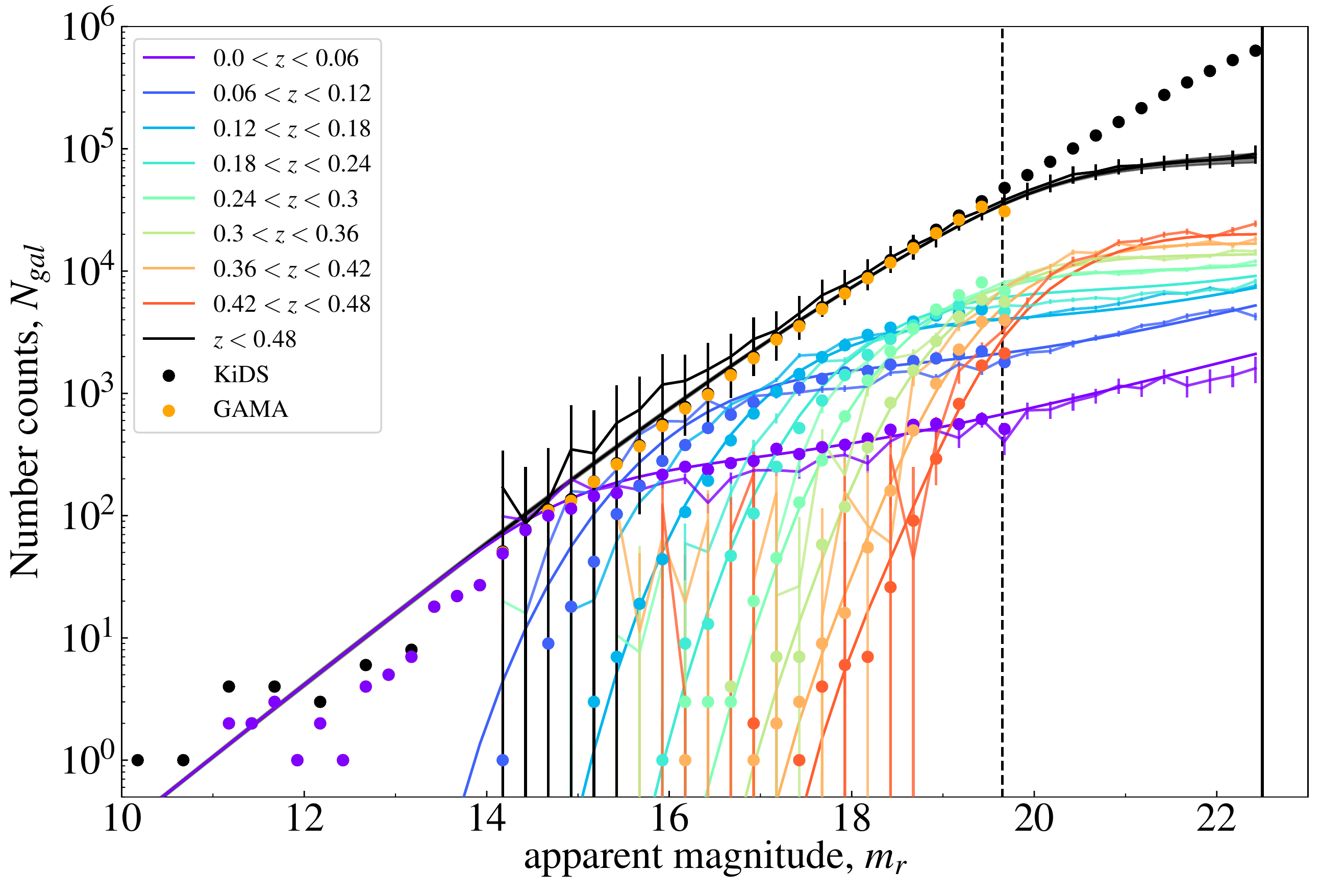}
	\caption{Number counts of the calculated model (solid lines) compared with the A-factor normalised cluster-$z$s (line with error bars) and the measured GAMA values (solid points), highlighting a series of redshift shells with corresponding normalised cluster-$z$s. The build-up of the total galaxy population and its agreement with the GAMA spec-$z$s where available can be seen. The decrease of the slope at fainter magnitudes results from our lack of information at redshifts $z>0.48$.}
	\label{fig:LFJarrett}
\end{figure*}

Having investigated the individual bins, we are now in the position to combine the data in order to determine the shape of the complete magnitude functions. In Fig. \ref{fig:LFJarrett} it can be seen the extent to which the A-factor normalised cluster-$z$s are in agreement with the GAMA measurements at $m_r<19.65$. At $m_r=19.65$ the total number counts of the model and our results summed over all redshift bins is flattening in contrast to the KiDS number counts. This gap between the models/results and the total number counts is explained by the increasing proportion of $z > 0.48$ population. Here no cluster-$z$s can be derived, due to the limitations of the reference sample. In Fig. \ref{fig:LFJarrett} a series of redshift shells is also displayed, and the model as well as the cluster-$z$s in the corresponding redshift shell is shown. The shape of the GLFs consistently displays the characteristic upturn of the Schechter function at bright magnitudes, followed by a flattening in the slope. In the low-z shells the slope is stable and almost linear over the whole range of magnitudes. For the highest redshift shells only the bright end of the GLF can be shown, and any information about the faint end is beyond the redshift range of this study. In summary, in all redshift shells the behaviour of the GLF is similar, and the number of galaxies continues to increase with fainter magnitudes, with slightly increasing slope.

\subsection{Measurement of the faint end GLF}
\label{sec:faint-end}
\begin{figure*}
	\centering
	\includegraphics[width=2\columnwidth]{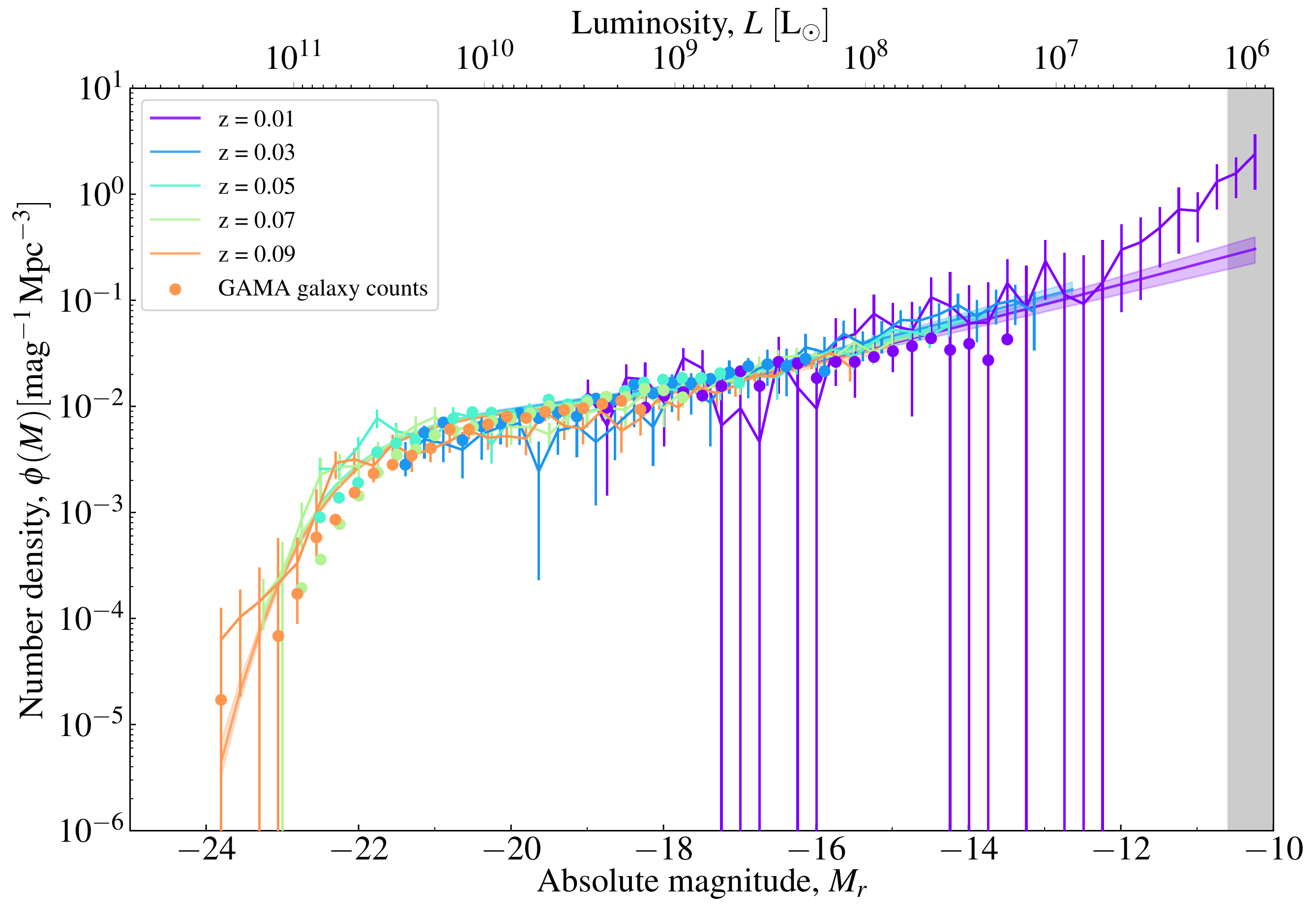}%
	\caption{Density distribution of the luminosity function at $z<0.1$. Here the resulting model (solid line) is compared to the A-factor normalised cluster-$z$s (line with error bars) and the measured GAMA values (solid points). The area in grey shows points which are potentially incomplete. An additional steepening of the GLF at the faint end is observed.}
	\label{fig:LFZoom} 
\end{figure*}

We now focus on the low-z ($z<0.1$) GLF itself, which is shown in Fig. \ref{fig:LFZoom}. We note that measurements of the GLF without the cluster-$z$s would only be possible up to $M_r = -13.5$; $L \sim 10^7 L_\odot$ (assuming $M_\odot = 4.65$). The cluster-$z$s provide almost $3$ additional magnitudes of information reaching down to $M_r = -10.7$ or $L \sim 10^6 L_\odot$. The GLF can hence be constructed over a total range of almost $14$ magnitudes.

In Fig. \ref{fig:LFZoom} it can be seen how the combination of the different redshift slices at $z<0.1$ are contributing and collectively building the GLF. The overlap of the cluster-$z$s, shown as lines with error bars, and the dots representing the spec-$z$s, are always in good agreement. Also the model is in agreement with the spec-$z$s and the cluster-$z$s. The shape of the GLF at the brightest magnitudes of $M_r \lesssim -20$ represents the characteristic cut-off of the GLF. Due to small volumes, this cut-off is only visible at redshifts $z>0.06$. With fainter magnitudes the GLF is extended successively by measurements of lower redshift bins, until it is unfolded over its full range. After the steep increase, the GLF flattens around $M_r \sim -20$ for a limited range. At fainter magnitudes the contribution of the second Schechter function becomes dominant, resulting in a slight increase in slope from $M_r \gtrsim -17$. This behaviour is not only true for the GAMA spec-$z$s, but also for the cluster-$z$s. As this trend remains unbroken until the limits of our study at $M_r = -10.7$, we conclude that the integral of the GLF (i.e. the number of galaxies in the Universe) remains divergent.

\section{Discussion}
\label{sec:Discussion}

\subsection{Modelling versus Measurement}
\label{modelvsmeasurement}
In this study we have derived results in two forms: firstly the best fit parametric model, which has been conditioned on both the spectroscopic and cluster-$z$ measurements, and secondly the observed GLF as derived from the model-normalised cluster-$z$ redshift distributions. As it can be seen in Fig. \ref{fig:LFZoom}, the modelled and the derived GLF measurements diverge for the faintest magnitudes, as the slope of the measurements is steeper than the best-fitting model. The question arises as to how to understand the nature of this discrepancy, and which determination ought to be preferred.

Since the model results necessarily depend on the choice of model parameterisation, this is an obvious first concern. Many different parameterisations are used in the literature, and we have no strong astrophysical justification for our particular choice. We explore the impact of model choice in Appendix \ref{app:choice of model}, where we use a simpler single Schecter model for the evolving GLF. Unsurprisingly, the resultant fit is quite different at the faint end, which is generally less well constrained by the data than around $M^*$.

What is more surprising is that although the model is rather different, the model-derived values for the normalisation constants $A_m$ are very robust.  
As shown in Fig.\ \ref{fig:SinglevsDouble}, the derived GLF measurements are hardly changed when we use this much simpler model.
In light of this fact, we prefer to view the parametric model mainly as a tool to derive the critical normalisation factors, by providing a self-consistent description of the full cluster-$z$ dataset, and we choose to focus instead on the model-normalised cluster-$z$ results as providing the more robust measurements of the evolving GLF.

\subsection{The steepening slope of the galaxy luminosity function at $z \sim 0$}
\label{sec:SteepeningSlope}

\begin{figure}
	\centering
	\includegraphics[width=\columnwidth]{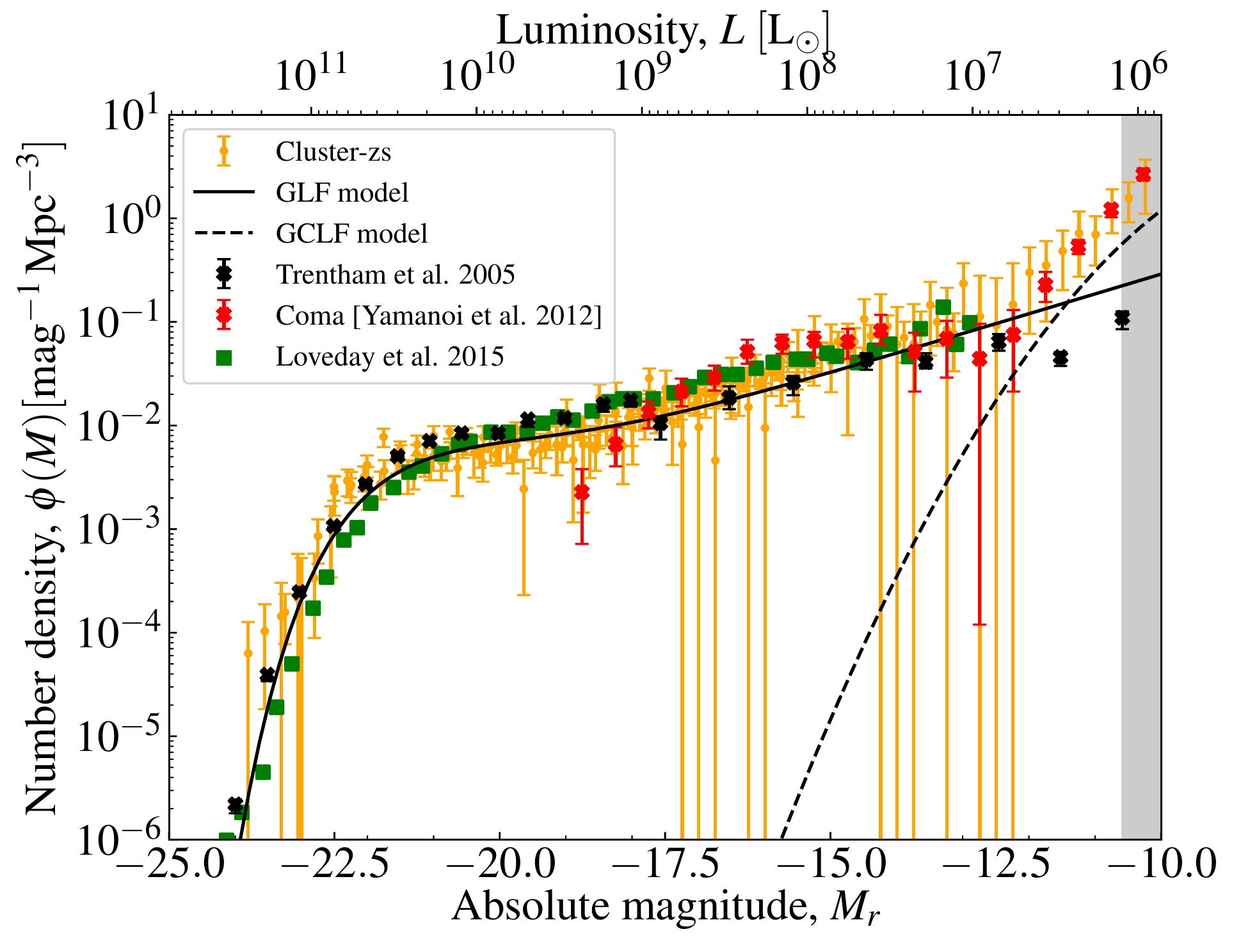}%
	\caption{Comparison of the resulting GLF measurements (orange) with the literature and the GLF and GCLF models (black). The values for the Coma Cluster are scaled by eye and the values $M_r>-10.7$ (grey area) are below our magnitude limits and potentially incomplete. At magnitudes $M_r\gtrsim-11.5$ the GCLF is resulting in larger values than the GLF.}
	\label{fig:LFComparison} 
\end{figure}

\begin{figure}
	\centering
	\includegraphics[width=\columnwidth]{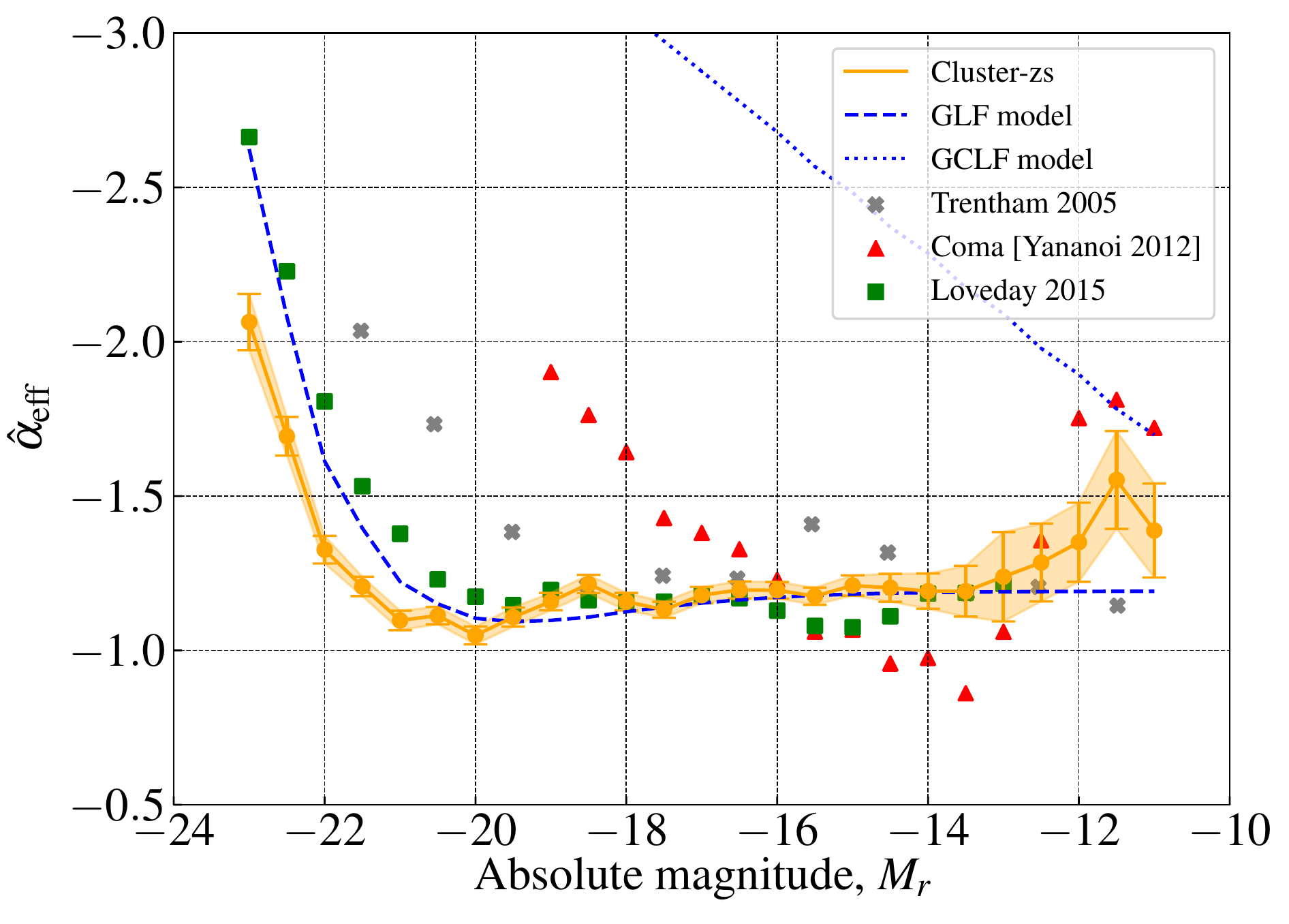}%
	\caption{Effective slope measured by fitting a power-law within overlapping bins of $\Delta mag=2$. We compare the slope of the cluster-$z$s (orange) to the GLF and GCLF models (blue) as well as studies by \protect\cite{Trentham2005} (grey) and \protect\cite{Yamanoi2012} (red). The measured effective slope is diverging from the GLF slope at fainter magnitudes towards the GCLF slope and is settling in between the predicted slopes of the GLF and GCLF at the faint end.}
	\label{fig:AvsMag} 
\end{figure}

One primary motivation for this study was to measure the shape of the GLF at the very faintest luminosities. In Fig.\ \ref{fig:LFComparison} we compare our GLF measurements to selected literature results. To directly compare the inferred shape of the GLF as observed by different studies, in Fig.\ \ref{fig:AvsMag} we also show the effective GLF slope, averaging over bins of width 2 mag.

For $-20 \lesssim M_r \lesssim -13$, we see good agreement between our measurements, \cite{Loveday2015}, \cite{Trentham2005} and the GLF model, with a nearly constant slope $\hat{\alpha}_{eff} \approx -1.2$. At fainter magnitudes we see a significant upturn in the cluster-$z$s measurements for $\log (L/L_\odot) \lesssim 6.5$, which is not captured by our parametric model for the GLF. While \cite{Trentham2005} does not see a similar upturn for field galaxies, a similarly steep upturn has been measured over the same magnitude range by \cite{Yamanoi2012} in the Coma Cluster. For Coma, \cite{Yamanoi2012} concluded that at $M_r > -12 $ GCs make up as much as $15\%$ of the total population. 

One possible explanation is thus that we are seeing globular clusters (GCs) and/or ultra compact dwarfs (UCDs) come to dominate the extragalactic source population in the field at these very low luminosities. To test this idea, we use our parametric GLF model to make a simple estimate for the expected GC luminosity function, as follows. We obtain the mean number of GCs as a function of magnitude 
\begin{align}
N_{GC} = S_n 10^{(-0.4 \times (M_V + 15) )}
\end{align}
\citep[][]{Harris1981}, where $S_n$ is the specific frequency of GCs normalized to a galaxy with an absolute magnitude of $M_V=-15$.  For the conversion between $r$- and $V$-band magnitudes we use $M_V = M_r + 0.25 \pm 0.07\rm{mag}$ , which we derived from the SED fits described by \cite{Taylor2011}. The shape of the GCLF is described by a Gaussian distribution function \citep[][]{Harris1991}:
\begin{align}
    N_{GC}(m) \propto e^{-(m-m_0)^2/2\sigma^2} ~ ,
\end{align}
described by the turnover magnitude of the distribution,  $m_0$, and the dispersion,  $\sigma$. The particular values we use for the derivation of the GCLF are based on studies of the Coma Cluster, where $m_0 = 27.7$, $\sigma = 1.48$ \citep[][]{Harris2009} and $S_n = 5.1$ \citep[][]{Franch2002}. The net cosmic-averaged GCLF is then obtained by integrating over our double Schecter model for the $z \sim 0$ GLF. 

This simple model, shown as the dashed line in  Fig. \ref{fig:LFComparison}, can be seen to do a remarkably good job at reproducing the steepening slope that we see for $\log (L/L_\odot) \lesssim 6.5$, which is where we expect the GC/UCD population to begin to outnumber the galaxy population. While we cannot distinguish between GCs and UCDs for our sample, we note that \cite{Mieske2012} argues that, at least in terms of luminosity distribution, UCDs can be viewed as continuing the bright tail of the Gaussian GC population. Thus we would appear to have mapped the GLF all the way down to the point where sub-galactic objects (i.e. GCs and UCDs) come to dominate in the field.

\subsection{What doesn't matter: errors/uncertainties that have little to no impact on our results.}
\label{sec:DoesntMatter}
In Sec.\ \ref{modelvsmeasurement}, we have already addressed the insensitivity of our results to the choice of parameterisation for the GLF model. Below we briefly describe several other tests we have performed to demonstrate the robustness of our analysis and results.

One potential concern is that the inclusion of stars, false detections or other 'bad' data will propagate through to bias the clustering-based redshift inferences. As described in Sec.\ \ref{sec:data}, we have excluded all entries in the photometric catalogue that are classified as either stars or artefacts, and only considered those classified as galaxies or ambiguous. However, the exclusion of stars and point-like objects such as higher-z galaxies (including QSOs) and even false detections is not necessary for deriving the cluster-$z$s. This is because stars do not cluster in the same way as galaxies, and so do not contribute to the cross-correlation function that is used to derive the cluster-$z$s. The same is true for artefacts, QSOs, and any other source population that does not follow the same large scale structure, as traced by the reference sample. The only effect would be a general dilution of the resulting clustering amplitude, which is accounted for by the $A_m$ and so do not influence the resulting GLF measurement.

Another possible concern stems from our use of the spec-$z$ sample to constrain the overall normalisation of the GLF via the characteristic density, $\phi^*$. How do we know that our results are not being driven by the spec-$z$ constraints rather the cluster-$z$s? We have addressed this concern by only using a bright ($m_r < 17.8$) subset of the spec-$z$ sample for our GLF model fitting, and verifying that we obtained similar results. 

An additional potential source of error is field-to-field variations. By calculating the cluster-$z$s for each field individually, and treating each of them equally during the fit, we are able to minimize the error as the $A_m$s account for variations between the three regions. These variations can be seen in Fig. \ref{fig:Afac}. Using this approach we are able to obtain the best results by combining the resulting measurements of each region into our final GLF.

The primary source of incompleteness is likely to be tied to low surface brightness, which will translate directly into an underestimate of the cluster-$z$ derived $P(z)$.
What matters is what proportion of the population we are missing. In light of the fact that apparent SB diminishes as $(1+z)^4$, it is an open question as to whether SB incompleteness will be a bigger issue for intrinsically fainter galaxies at low redshift, or for much more numerous higher redshift galaxies. 

If it is the former, then this will lead us to underestimate the GLF for the faintest galaxies at $z \sim 0$, and our measurements should be taken as a lower limit. In Sec \ref{sec:limits}, we describe how we have limited our analysis to $m_r < 22.55$ to minimise the impact of SB selection effects, and particularly incompleteness for low SB galaxies. Based on Fig.\ \ref{fig:Surface-Brightness}, we can estimate an approximate SB selection limit around $\mu \sim 26 \rm{mag/arcsec}^2$: that is, faint enough to capture even the extreme population of Ultra Diffuse Galaxies (UDGs) found by \citet{Dokkum2015} to redshifts less than $\sim 0.1$.

If surface brightness selection effects become increasingly important for higher redshifts, then the impact on our results is less clear. What will happen is that our measured $P(z)$s will be systematically low for the highest redshifts and faintest magnitudes. In principle, this might lead to an overestimate of the $A_m$ normalisation factors for the faintest magnitudes, and so lead to a steepening of the observed GLF slopes at all redshifts. What makes this difficult to predict is not knowing how the modelling might respond to this systematic change in the data. While we cannot exclude this possibility, we do see good agreement in the cluster-$z$ derived GLF measurements across different $z$ ranges (see Fig.\ \ref{fig:LFZoom}), which suggests that the impact of this kind of effect is small.

With these considerations, we can conclude that our approach is insensitive to many difficulties in deriving the GLF.

\subsection{What does matter: bias evolution is the limiting source of error/uncertainty}
\label{sec:DoesMatter}
The main source of systematic uncertainty in our study is tied to the unknown differential galaxy bias evolution $\overline{b}_t(m,z)$ of the target dataset. In general the form of the bias can be measured where spec-$z$s are available, and the corresponding bias of the reference dataset is accounted for by calculating the auto-correlation function in each redshift bin \citep[see][]{Busch2020}. However, this is impossible for the target dataset. We therefore have to address any effects due to the magnitude and redshift dependence of the target galaxy bias. 

Any magnitude-dependent bias would result in a degeneracy with the density evolution of $\phi_0^*$ and therefore  the shape of the GLF. As we are working in apparent magnitude bins, galaxies across a broad redshift range, as well as with different luminosities, are included in one bin. As the bias should be larger for brighter galaxies and smaller at higher redshift, these effects might cancel out to some extent. 

Bias dependence with magnitude is also partially accounted for by the A-factors, as any change of normalisation of the cluster-$z$s in each magnitude would impact the $A_m$s, but not the final measurements. Even under the assumption of a constant $A \times N$, by which the cluster-$z$s are normalised only based onto the number of objects in each magnitude bin and hence ignoring the bias completely, we get a sensible faint end slope of $\alpha \sim -1.6$, which shows that the magnitude evolution of the unknown bias is rather small.

The bias evolution in redshift is of larger concern. In contrast to the magnitude bias, the effect of the redshift-dependent bias is not to change the shape of the GLF, but its evolution. The impact of the redshift bias can be seen in Fig. \ref{fig:clusterzs} and Fig. \ref{fig:Nz}. Corrections to the linear bias of the form $\delta \bar{b}_t/\delta z = 1$ are suggested by \cite{Rahman2015} and \cite{Bates2019}. In an approach by \cite{vanDaalen2018} it is suggested that by using a simple luminosity bias relation with a fixed and known normalisation, the redshift evolution of the remaining bias terms cancel out. We have performed tests which show that corrections using the shape of a power law $\bar{b}_t(z) = (1+z)^\beta$, with $\beta \approx 1$, as shown by \cite{Davis2018}, can determine the known distribution of the GAMA spec-$z$s in agreement with the inferred cluster-$z$s. Unfortunately all of these bias corrections can only be tested where spec-$z$s are available. Even though there are good reasons for the use of a bias correction, for  reasons of simplicity we have chosen to use a constant $\bar{b}_t$. In addition, as we focus on the low-z GLF ($z<0.1$), the effect of an uncorrected bias is unimportant for our main conclusion. Nevertheless, the unknown bias remains the main systematic uncertainty in this study.

\section{Summary and Conclusion}
\label{sec:Summary}
In this paper, we have demonstrated a novel experimental design for using clustering redshifts to measure the evolving GLF, and especially the GLF shape at $z \sim 0$, to the faintest luminosities, beyond the limits of spec-$z$s and notably beyond the useful limits of photometric redshifts.

Our GLF final measurements are based on a sample of $\sim3\times10^6$ sources to $m_r < 22.5$ (i.e. $\sim3$ magnitudes beyond the GAMA spectroscopic redshift limit), using the three GAMA equatorial fields. Our experiment considers {\em only} position and total $r$-band magnitude for this sample. The information is taken from the GAMA produced photometric catalogues \citep{KiDSBellstedt2020}, which are derived from KiDS $r$-band imaging \citep{KiDSKuijken2019}. 

As discussed in Sec. \ref{sec:limits}, we have limited our analysis to $m_r < 22.5$ to minimise the impact of SB selection effects on our results (see Fig.\ \ref{fig:Surface-Brightness}).

As illustrated in Fig.\ \ref{fig:Cluster-Explanation}, our clustering redshift inferences are based on the cross-correlation between the target sample and a reference sample with known redshifts, where  the size of the reference sample limits the statistical precision of our experiment, and also the maximum redshift interval that we can probe. We use the main GAMA spectroscopic redshift sample ($m_r < 19.65$; $z \lesssim 0.5$; $N \sim 170.000$) for this purpose. 

In Sec. \ref{sec:recovery-number-dist} we have demonstrated that we can use clustering redshift inference to recover $N(z_\mathrm{spec}|m)$ --- the spectroscopic redshift distribution in bins of apparent magnitude --- for the GAMA sample (see Fig.\ \ref{fig:Nz}).

The main technical challenge in our experiment arises from the fact that output of the process of clustering redshift inference is proportional to the redshift distribution for the target sample, up to some unknown scalar (see Sec. \ref{sec:Afac}). Our strategy is to use a simple parametric model for the evolving GLF to constrain the values for the normalisation factors, $A_m$, as described in Sec. \ref{sec:Afac} (see also Fig.\ \ref{fig:Afac}). 

Fig. \ref{fig:AfacDist} provides an overview for how we use the results cluster-$z$ results to measure the GLF. We use clustering redshift inference to derive the redshift distribution for our target sample in bins of apparent magnitude, $P(z|m)$. The derived values of the normalisation factors, $A_m$, then are used to obtain the number counts, $N(z|m)$ (see Fig.s \ref{fig:Nz} and \ref{fig:LFJarrett}). Finally for a given cosmology to determine the distance modulus and $dV/dz$, the luminosity function $\Phi(M|z)$ follows.

Our main results --- mapping the field GLF at $z \sim 0$ across $14$ magnitudes or $5.5$ decades in luminosity -- are shown in Fig. \ref{fig:LFZoom} and Fig. \ref{fig:LFComparison}. The measured slope of the GLF remains remarkably flat over the range $-20 \lesssim M_r \lesssim 13$, with a sharp upturn below $M_r \sim -12.5$ or $\log {L/L_\odot} \sim 6.5$. A similar upturn has been found for the Coma Cluster by \citet{Yamanoi2012}. Following \citet{Yamanoi2012}, we use a simple model to predict the luminosity function for the GC population, based on our GLF fits. This simple prediction with no free parameters provides a good explanation to the observations.

As discussed in Sec. \ref{sec:DoesntMatter}, we have conducted a number of sensitivity tests to demonstrate that our results are robust to a variety of elements of the experimental design, including: model parameterisation; the presence of stars, QSOs, artefacts, etc.\ in the photometric catalog; and the depth of the spec-$z$ sample used to constrain the overall GLF normalisation, $\phi^*$. Also potential effects due to SB selection were discussed supplementary to our measures to minimise its impact.

The dominant source of systematic error/uncertainty in our results is the unknown evolution of the mean bias of the target samples over the $0 < z \lesssim 0.5$ interval. Being mindful of these issues, we have focused particularly on the $z \sim 0$ GLF, where the impact of these uncertainties is minimised.

Thus we have mapped the $z\sim0$ GLF from the most luminous galaxies all the way down to where sub-galactic objects like GCs and UCDs take over as the most numerous extragalactic population.

In doing so we demonstrated the potential for clustering based redshift inference in deriving the GLF. This technique offers manifold applications as it is not limited to the optical only. In addition this technique can be extended: e.g. by using deeper reference sets, or by combining different reference sets, an even deeper study would be possible.

\section*{Data availability}
The  derived  GLF  measurements  shown  in  this  article  are available in the article and in its online supplementary material.

\section*{Acknowledgements}
GSK acknowledges financial support received through a Swinburne University Postgraduate Research Award. GAMA is an European-Australasian spectroscopic survey using the Anglo-Australian Telescope. Its input catalogue is based on data taken from the Sloan Digital Sky Survey and the UKIRT Infrared Deep Sky Survey. In addition complementary imaging is being obtained by a number of independent survey programs including GALEX MIS, VST KIDS, VISTA VIKING, WISE, Herschel-ATLAS, GMRT and ASKAP providing UV to radio coverage. GAMA is funded by the STFC (UK), the ARC (Australia), the AAO, and the participating institutions. The GAMA website is http://www.gama-survey.org/. This work is on based on observations made with ESO Telescopes at the La Silla Paranal Observatory under programme IDs 177.A-3016, 177.A-3017, 177.A-3018 and 179.A-2004, and on data products produced by the KiDS consortium. The KiDS production team acknowledges support from: Deutsche Forschungsgemeinschaft, ERC, NOVA and NWO-M grants; Target; the University of Padova, and the University Federico II (Naples). HH is supported by a Heisenberg grant of the Deutsche Forschungsgemeinschaft (Hi 1495/5-1). HH and AHM are supported by an ERC Consolidator Grant (No. 770935). MB is supported by the Polish National Science Center through grants no. 2020/38/E/ST9/00395, 2018/30/E/ST9/00698 and 2018/31/G/ST9/03388, and by the Polish Ministry of Science and Higher Education through grant DIR/WK/2018/12.

For this study \texttt{PYTHON} has been used for the data analysis, and we acknowledge the use of Matplotlib \citep[][]{Hunter2007} for the generation of figures in this paper.
%%%%%%%%%%%%%%%%%%% REFERENCES %%%%%%%%%%%%%%%%%%
\balance
\bibliographystyle{aa}
\bibliography{sources}

\begin{thebibliography}{84}
\expandafter\ifx\csname natexlab\endcsname\relax\def\natexlab#1{#1}\fi

\bibitem[{{Baldry} {et~al.}(2012){Baldry}, {Driver}, {Loveday}, {Taylor},
  {Kelvin}, {Liske}, {Norberg}, {Robotham}, {Brough}, {Hopkins}, {Bamford},
  {Peacock}, {Bland-Hawthorn}, {Conselice}, {Croom}, {Jones}, {Parkinson},
  {Popescu}, {Prescott}, {Sharp}, \& {Tuffs}}]{Baldry2012}
{Baldry}, I.~K., {Driver}, S.~P., {Loveday}, J., {et~al.} 2012, \mnras, 421,
  621

\bibitem[{{Baldry} {et~al.}(2008){Baldry}, {Glazebrook}, \&
  {Driver}}]{Baldry2008}
{Baldry}, I.~K., {Glazebrook}, K., \& {Driver}, S.~P. 2008, \mnras, 388, 945

\bibitem[{{Bates} {et~al.}(2019){Bates}, {Tojeiro}, {Newman}, {Gonzalez-Perez},
  {Comparat}, {Schneider}, {Lima}, \& {Streblyanska}}]{Bates2019}
{Bates}, D.~J., {Tojeiro}, R., {Newman}, J.~A., {et~al.} 2019, \mnras, 486,
  3059

\bibitem[{{Bellstedt} {et~al.}(2020){Bellstedt}, {Driver}, {Robotham},
  {Davies}, {Bogue}, {Cook}, {Hashemizadeh}, {Koushan}, {Taylor}, {Thorne},
  {Turner}, \& {Wright}}]{KiDSBellstedt2020}
{Bellstedt}, S., {Driver}, S.~P., {Robotham}, A. S.~G., {et~al.} 2020, \mnras,
  496, 3235

\bibitem[{{Bertin} \& {Arnouts}(1996)}]{BertinArnouts1996}
{Bertin}, E. \& {Arnouts}, S. 1996, \aaps, 117, 393

\bibitem[{{Blanton} {et~al.}(2005){Blanton}, {Lupton}, {Schlegel}, {Strauss},
  {Brinkmann}, {Fukugita}, \& {Loveday}}]{Blanton2005}
{Blanton}, M.~R., {Lupton}, R.~H., {Schlegel}, D.~J., {et~al.} 2005, \apj, 631,
  208

\bibitem[{{Bower} {et~al.}(2012){Bower}, {Benson}, \& {Crain}}]{Bower2012}
{Bower}, R.~G., {Benson}, A.~J., \& {Crain}, R.~A. 2012, \mnras, 422, 2816

\bibitem[{{Chiboucas} {et~al.}(2009){Chiboucas}, {Karachentsev}, \&
  {Tully}}]{Chiboucas2009}
{Chiboucas}, K., {Karachentsev}, I.~D., \& {Tully}, R.~B. 2009, \aj, 137, 3009

\bibitem[{{Choi} {et~al.}(2016){Choi}, {Heymans}, {Blake}, {Hildebrandt},
  {Duncan}, {Erben}, {Nakajima}, {Van Waerbeke}, \& {Viola}}]{Choi2016}
{Choi}, A., {Heymans}, C., {Blake}, C., {et~al.} 2016, Monthly Notices of the
  Royal Astronomical Society, 463, 3737

\bibitem[{{Cole} {et~al.}(1994){Cole}, {Aragon-Salamanca}, {Frenk}, {Navarro},
  \& {Zepf}}]{Cole1994}
{Cole}, S., {Aragon-Salamanca}, A., {Frenk}, C.~S., {Navarro}, J.~F., \&
  {Zepf}, S.~E. 1994, \mnras, 271, 781

\bibitem[{{Cole} {et~al.}(2005){Cole}, {Percival}, {Peacock}, {Norberg},
  {Baugh}, {Frenk}, {Baldry}, {Bland-Hawthorn}, {Bridges}, {Cannon}, {Colless},
  {Collins}, {Couch}, {Cross}, {Dalton}, {Eke}, {De Propris}, {Driver},
  {Efstathiou}, {Ellis}, {Glazebrook}, {Jackson}, {Jenkins}, {Lahav}, {Lewis},
  {Lumsden}, {Maddox}, {Madgwick}, {Peterson}, {Sutherland}, \&
  {Taylor}}]{Cole2005}
{Cole}, S., {Percival}, W.~J., {Peacock}, J.~A., {et~al.} 2005, \mnras, 362,
  505

\bibitem[{{Cross} \& {Driver}(2002)}]{Cross2002}
{Cross}, N. \& {Driver}, S.~P. 2002, \mnras, 329, 579

\bibitem[{{Cross} {et~al.}(2001){Cross}, {Driver}, {Couch}, {Baugh},
  {Bland-Hawthorn}, {Bridges}, {Cannon}, {Cole}, {Colless}, {Collins},
  {Dalton}, {Deeley}, {De Propris}, {Efstathiou}, {Ellis}, {Frenk},
  {Glazebrook}, {Jackson}, {Lahav}, {Lewis}, {Lumsden}, {Maddox}, {Madgwick},
  {Moody}, {Norberg}, {Peacock}, {Peterson}, {Price}, {Seaborne}, {Sutherland},
  {Tadros}, \& {Taylor}}]{Cross2001}
{Cross}, N., {Driver}, S.~P., {Couch}, W., {et~al.} 2001, \mnras, 324, 825

\bibitem[{{Croton} {et~al.}(2006){Croton}, {Springel}, {White}, {De Lucia},
  {Frenk}, {Gao}, {Jenkins}, {Kauffmann}, {Navarro}, \& {Yoshida}}]{Croton2006}
{Croton}, D.~J., {Springel}, V., {White}, S. D.~M., {et~al.} 2006, \mnras, 365,
  11

\bibitem[{{Davis} {et~al.}(2018){Davis}, {Rozo}, {Roodman}, {Alarcon},
  {Cawthon}, {Gatti}, {Lin}, {Miquel}, {Rykoff}, {Troxel}, {Vielzeuf},
  {Abbott}, {Abdalla}, {Allam}, {Annis}, {Bechtol}, {Benoit-L{\'e}vy},
  {Bertin}, {Brooks}, {Buckley-Geer}, {Burke}, {Carnero Rosell}, {Carrasco
  Kind}, {Carretero}, {Castander}, {Crocce}, {Cunha}, {D'Andrea}, {da Costa},
  {Desai}, {Diehl}, {Doel}, {Drlica-Wagner}, {Fausti Neto}, {Flaugher},
  {Fosalba}, {Frieman}, {Garc{\'\i}a-Bellido}, {Gaztanaga}, {Gerdes},
  {Giannantonio}, {Gruen}, {Gruendl}, {Gutierrez}, {Honscheid}, {Jain},
  {James}, {Jeltema}, {Krause}, {Kuehn}, {Kuhlmann}, {Kuropatkin}, {Lahav},
  {Li}, {Lima}, {March}, {Marshall}, {Martini}, {Melchior}, {Ogando}, {Plazas},
  {Romer}, {Sanchez}, {Scarpine}, {Schindler}, {Schubnell}, {Sevilla-Noarbe},
  {Smith}, {Soares-Santos}, {Sobreira}, {Suchyta}, {Swanson}, {Tarle},
  {Thomas}, {Vikram}, {Walker}, {Wechsler}, \& {DES Collaboration}}]{Davis2018}
{Davis}, C., {Rozo}, E., {Roodman}, A., {et~al.} 2018, \mnras, 477, 2196

\bibitem[{{Dawson} {et~al.}(2013){Dawson}, {Schlegel}, {Ahn}, {Anderson},
  {Aubourg}, {Bailey}, {Barkhouser}, {Bautista}, {Beifiori}, {Berlind},
  {Bhardwaj}, {Bizyaev}, {Blake}, {Blanton}, {Blomqvist}, {Bolton}, {Borde},
  {Bovy}, {Brandt}, {Brewington}, {Brinkmann}, {Brown}, {Brownstein}, {Bundy},
  {Busca}, {Carithers}, {Carnero}, {Carr}, {Chen}, {Comparat}, {Connolly},
  {Cope}, {Croft}, {Cuesta}, {da Costa}, {Davenport}, {Delubac}, {de Putter},
  {Dhital}, {Ealet}, {Ebelke}, {Eisenstein}, {Escoffier}, {Fan}, {Filiz Ak},
  {Finley}, {Font-Ribera}, {G{\'e}nova-Santos}, {Gunn}, {Guo}, {Haggard},
  {Hall}, {Hamilton}, {Harris}, {Harris}, {Ho}, {Hogg}, {Holder}, {Honscheid},
  {Huehnerhoff}, {Jordan}, {Jordan}, {Kauffmann}, {Kazin}, {Kirkby}, {Klaene},
  {Kneib}, {Le Goff}, {Lee}, {Long}, {Loomis}, {Lundgren}, {Lupton}, {Maia},
  {Makler}, {Malanushenko}, {Malanushenko}, {Mandelbaum}, {Manera}, {Maraston},
  {Margala}, {Masters}, {McBride}, {McDonald}, {McGreer}, {McMahon}, {Mena},
  {Miralda-Escud{\'e}}, {Montero-Dorta}, {Montesano}, {Muna}, {Myers},
  {Naugle}, {Nichol}, {Noterdaeme}, {Nuza}, {Olmstead}, {Oravetz}, {Oravetz},
  {Owen}, {Padmanabhan}, {Palanque-Delabrouille}, {Pan}, {Parejko},
  {P{\^a}ris}, {Percival}, {P{\'e}rez-Fournon}, {P{\'e}rez-R{\`a}fols},
  {Petitjean}, {Pfaffenberger}, {Pforr}, {Pieri}, {Prada}, {Price-Whelan},
  {Raddick}, {Rebolo}, {Rich}, {Richards}, {Rockosi}, {Roe}, {Ross}, {Ross},
  {Rossi}, {Rubi{\~n}o-Martin}, {Samushia}, {S{\'a}nchez}, {Sayres}, {Schmidt},
  {Schneider}, {Sc{\'o}ccola}, {Seo}, {Shelden}, {Sheldon}, {Shen}, {Shu},
  {Slosar}, {Smee}, {Snedden}, {Stauffer}, {Steele}, {Strauss}, {Streblyanska},
  {Suzuki}, {Swanson}, {Tal}, {Tanaka}, {Thomas}, {Tinker}, {Tojeiro},
  {Tremonti}, {Vargas Maga{\~n}a}, {Verde}, {Viel}, {Wake}, {Watson}, {Weaver},
  {Weinberg}, {Weiner}, {West}, {White}, {Wood-Vasey}, {Yeche}, {Zehavi},
  {Zhao}, \& {Zheng}}]{BOSS2013}
{Dawson}, K.~S., {Schlegel}, D.~J., {Ahn}, C.~P., {et~al.} 2013, \aj, 145, 10

\bibitem[{{Dekel} \& {Silk}(1986)}]{Dekel1986}
{Dekel}, A. \& {Silk}, J. 1986, \apj, 303, 39

\bibitem[{{Driver} {et~al.}(2021){Driver}, {Bellstedt}, {Robotham}, {Baldry},
  {Davies}, {Liske}, {Obreschkow}, {Taylor}, {Wright}, {Alpaslan}, {Bamford},
  {Bauer}, {Bland-Hawhorn}, {Cluver}, {Colless}, J., {Croom}, {de Jong}, {De
  Propis}, {Drinkwater}, {Dvornik}, {Farrow}, {Frenk}, {Giblin}, {Graham},
  {Grootes}, {Gunawardhana}, {Hashemizadeh}, {Häusler}, {Heymans},
  {Hildebrandt}, {Holwerda}, {Hopkins}, {Jarrett}, {Jones}, {Kelvin},
  {Koushan}, {Kuijken}, {Lara-Lopez}, {Lange}, {Lopez-Sanchez}, {Loveday},
  {Mahajan}, {Moffett}, {Napolitano}, {Norberg}, {Owers}, {Radovich}, {Raouf},
  {Peacock}, {Phillips}, {Pimblett}, {Popescu}, {Sansom}, {Seibert},
  {Sutherland}, {Thorne}, {Tuffis}, {Turner}, {van Kampen}, \&
  {Wilkins}}]{Driver2021}
{Driver}, S.~P., {Bellstedt}, S., {Robotham}, A.~S., {et~al.} 2021, \mnras

\bibitem[{{Driver} \& {Phillipps}(1996)}]{Driver1996}
{Driver}, S.~P. \& {Phillipps}, S. 1996, \apj, 469, 529

\bibitem[{{Driver} {et~al.}(1994){Driver}, {Phillipps}, {Davies}, {Morgan}, \&
  {Disney}}]{Driver1994}
{Driver}, S.~P., {Phillipps}, S., {Davies}, J.~I., {Morgan}, I., \& {Disney},
  M.~J. 1994, \mnras, 268, 393

\bibitem[{{Eisenstein} {et~al.}(2005){Eisenstein}, {Zehavi}, {Hogg},
  {Scoccimarro}, {Blanton}, {Nichol}, {Scranton}, {Seo}, {Tegmark}, {Zheng},
  {Anderson}, {Annis}, {Bahcall}, {Brinkmann}, {Burles}, {Castander},
  {Connolly}, {Csabai}, {Doi}, {Fukugita}, {Frieman}, {Glazebrook}, {Gunn},
  {Hendry}, {Hennessy}, {Ivezi{\'c}}, {Kent}, {Knapp}, {Lin}, {Loh}, {Lupton},
  {Margon}, {McKay}, {Meiksin}, {Munn}, {Pope}, {Richmond}, {Schlegel},
  {Schneider}, {Shimasaku}, {Stoughton}, {Strauss}, {SubbaRao}, {Szalay},
  {Szapudi}, {Tucker}, {Yanny}, \& {York}}]{Eisenstein2005}
{Eisenstein}, D.~J., {Zehavi}, I., {Hogg}, D.~W., {et~al.} 2005, \apj, 633, 560

\bibitem[{{Foreman-Mackey} {et~al.}(2019){Foreman-Mackey}, {Farr}, {Sinha},
  {Archibald}, {Hogg}, {Sanders}, {Zuntz}, {Williams}, {Nelson}, {de
  Val-Borro}, {Erhardt}, {Pashchenko}, \& {Pla}}]{emcee2019}
{Foreman-Mackey}, D., {Farr}, W., {Sinha}, M., {et~al.} 2019, The Journal of
  Open Source Software, 4, 1864

\bibitem[{{Gatti} {et~al.}(2018){Gatti}, {Vielzeuf}, {Davis}, {Cawthon}, {Rau},
  {DeRose}, {De Vicente}, {Alarcon}, {Rozo}, {Gaztanaga}, {Hoyle}, {Miquel},
  {Bernstein}, {Bonnett}, {Carnero Rosell}, {Castand er}, {Chang}, {da Costa},
  {Gruen}, {Gschwend}, {Hartley}, {Lin}, {MacCrann}, {Maia}, {Ogando},
  {Roodman}, {Sevilla-Noarbe}, {Troxel}, {Wechsler}, {Asorey}, {Davis},
  {Glazebrook}, {Hinton}, {Lewis}, {Lidman}, {Macaulay}, {M{\"o}ller},
  {O'Neill}, {Sommer}, {Uddin}, {Yuan}, {Zhang}, {Abbott}, {Allam}, {Annis},
  {Bechtol}, {Brooks}, {Burke}, {Carollo}, {Carrasco Kind}, {Carretero},
  {Cunha}, {D'Andrea}, {DePoy}, {Desai}, {Eifler}, {Evrard}, {Flaugher},
  {Fosalba}, {Frieman}, {Garc{\'\i}a-Bellido}, {Gerdes}, {Goldstein},
  {Gruendl}, {Gutierrez}, {Honscheid}, {Hoormann}, {Jain}, {James}, {Jarvis},
  {Jeltema}, {Johnson}, {Johnson}, {Krause}, {Kuehn}, {Kuhlmann}, {Kuropatkin},
  {Li}, {Lima}, {Marshall}, {Melchior}, {Menanteau}, {Nichol}, {Nord},
  {Plazas}, {Reil}, {Rykoff}, {Sako}, {Sanchez}, {Scarpine}, {Schubnell},
  {Sheldon}, {Smith}, {Smith}, {Soares-Santos}, {Sobreira}, {Suchyta},
  {Swanson}, {Tarle}, {Thomas}, {Tucker}, {Tucker}, {Vikram}, {Walker},
  {Weller}, {Wester}, \& {Wolf}}]{Gatti2018}
{Gatti}, M., {Vielzeuf}, P., {Davis}, C., {et~al.} 2018, Monthly Notices of the
  Royal Astronomical Society, 477, 1664

\bibitem[{{Goodman} \& {Weare}(2010)}]{Goodman2010}
{Goodman}, J. \& {Weare}, J. 2010, Communications in Applied Mathematics and
  Computational Science, 5, 65

\bibitem[{{Harris}(1991)}]{Harris1991}
{Harris}, W.~E. 1991, \araa, 29, 543

\bibitem[{{Harris} {et~al.}(2009){Harris}, {Kavelaars}, {Hanes}, {Pritchet}, \&
  {Baum}}]{Harris2009}
{Harris}, W.~E., {Kavelaars}, J.~J., {Hanes}, D.~A., {Pritchet}, C.~J., \&
  {Baum}, W.~A. 2009, \aj, 137, 3314

\bibitem[{{Harris} \& {van den Bergh}(1981)}]{Harris1981}
{Harris}, W.~E. \& {van den Bergh}, S. 1981, \aj, 86, 1627

\bibitem[{Hildebrandt {et~al.}(2016)Hildebrandt, Viola, Heymans, Joudaki,
  Kuijken, Blake, Erben, Joachimi, Klaes, Miller, \& et~al.}]{Hildebrandt2016}
Hildebrandt, H., Viola, M., Heymans, C., {et~al.} 2016, Monthly Notices of the
  Royal Astronomical Society, 465

\bibitem[{{Hunter}(2007)}]{Hunter2007}
{Hunter}, J.~D. 2007, Computing in Science and Engineering, 9, 90

\bibitem[{{Johnson} {et~al.}(2017){Johnson}, {Blake}, {Amon}, {Erben},
  {Glazebrook}, {Harnois-Deraps}, {Heymans}, {Hildebrandt}, {Joudaki}, {Klaes},
  {Kuijken}, {Lidman}, {Marin}, {McFarland}, {Morrison}, {Parkinson}, {Poole},
  {Radovich}, \& {Wolf}}]{Johnson2017}
{Johnson}, A., {Blake}, C., {Amon}, A., {et~al.} 2017, \mnras, 465, 4118

\bibitem[{{Kauffmann} {et~al.}(1993){Kauffmann}, {White}, \&
  {Guiderdoni}}]{Kauffmann1993}
{Kauffmann}, G., {White}, S.~D.~M., \& {Guiderdoni}, B. 1993, \mnras, 264, 201

\bibitem[{{Kelvin} {et~al.}(2014){Kelvin}, {Driver}, {Robotham}, {Graham},
  {Phillipps}, {Agius}, {Alpaslan}, {Baldry}, {Bamford}, {Bland-Hawthorn},
  {Brough}, {Brown}, {Colless}, {Conselice}, {Hopkins}, {Liske}, {Loveday},
  {Norberg}, {Pimbblet}, {Popescu}, {Prescott}, {Taylor}, \&
  {Tuffs}}]{Kelvin2014}
{Kelvin}, L.~S., {Driver}, S.~P., {Robotham}, A. S.~G., {et~al.} 2014, \mnras,
  439, 1245

\bibitem[{{Koposov} {et~al.}(2008){Koposov}, {Belokurov}, {Evans}, {Hewett},
  {Irwin}, {Gilmore}, {Zucker}, {Rix}, {Fellhauer}, {Bell}, \&
  {Glushkova}}]{Koposov2008}
{Koposov}, S., {Belokurov}, V., {Evans}, N.~W., {et~al.} 2008, \apj, 686, 279

\bibitem[{{Kuijken} {et~al.}(2019){Kuijken}, {Heymans}, {Dvornik},
  {Hildebrandt}, {de Jong}, {Wright}, {Erben}, {Bilicki}, {Giblin}, {Shan},
  {Getman}, {Grado}, {Hoekstra}, {Miller}, {Napolitano}, {Paolilo}, {Radovich},
  {Schneider}, {Sutherland}, {Tewes}, {Tortora}, {Valentijn}, \& {Verdoes
  Kleijn}}]{KiDSKuijken2019}
{Kuijken}, K., {Heymans}, C., {Dvornik}, A., {et~al.} 2019, \aap, 625, A2

\bibitem[{{Landy} \& {Szalay}(1993)}]{Landy1993}
{Landy}, S.~D. \& {Szalay}, A.~S. 1993, \apj, 412, 64

\bibitem[{{Lin} {et~al.}(1999){Lin}, {Yee}, {Carlberg}, {Morris}, {Sawicki},
  {Patton}, {Wirth}, \& {Shepherd}}]{Lin1999}
{Lin}, H., {Yee}, H.~K.~C., {Carlberg}, R.~G., {et~al.} 1999, \apj, 518, 533

\bibitem[{{Liske} {et~al.}(2015){Liske}, {Baldry}, {Driver}, {Tuffs},
  {Alpaslan}, {Andrae}, {Brough}, {Cluver}, {Grootes}, {Gunawardhana},
  {Kelvin}, {Loveday}, {Robotham}, {Taylor}, {Bamford}, {Bland-Hawthorn},
  {Brown}, {Drinkwater}, {Hopkins}, {Meyer}, {Norberg}, {Peacock}, {Agius},
  {Andrews}, {Bauer}, {Ching}, {Colless}, {Conselice}, {Croom}, {Davies}, {De
  Propris}, {Dunne}, {Eardley}, {Ellis}, {Foster}, {Frenk}, {H{\"a}u{\ss}ler},
  {Holwerda}, {Howlett}, {Ibarra}, {Jarvis}, {Jones}, {Kafle}, {Lacey},
  {Lange}, {Lara-L{\'o}pez}, {L{\'o}pez-S{\'a}nchez}, {Maddox}, {Madore},
  {McNaught-Roberts}, {Moffett}, {Nichol}, {Owers}, {Palamara}, {Penny},
  {Phillipps}, {Pimbblet}, {Popescu}, {Prescott}, {Proctor}, {Sadler},
  {Sansom}, {Seibert}, {Sharp}, {Sutherland}, {V{\'a}zquez-Mata}, {van Kampen},
  {Wilkins}, {Williams}, \& {Wright}}]{Liske2015}
{Liske}, J., {Baldry}, I.~K., {Driver}, S.~P., {et~al.} 2015, \mnras, 452, 2087

\bibitem[{{Loveday}(1997)}]{Loveday1997}
{Loveday}, J. 1997, \apj, 489, 29

\bibitem[{{Loveday} {et~al.}(2015){Loveday}, {Norberg}, {Baldry}, {Bland
  -Hawthorn}, {Brough}, {Brown}, {Driver}, {Kelvin}, \&
  {Phillipps}}]{Loveday2015}
{Loveday}, J., {Norberg}, P., {Baldry}, I.~K., {et~al.} 2015, \mnras, 451, 1540

\bibitem[{{Loveday} {et~al.}(2012){Loveday}, {Norberg}, {Baldry}, {Driver},
  {Hopkins}, {Peacock}, {Bamford}, {Liske}, {Bland-Hawthorn}, {Brough},
  {Brown}, {Cameron}, {Conselice}, {Croom}, {Frenk}, {Gunawardhana}, {Hill},
  {Jones}, {Kelvin}, {Kuijken}, {Nichol}, {Parkinson}, {Phillipps}, {Pimbblet},
  {Popescu}, {Prescott}, {Robotham}, {Sharp}, {Sutherland}, {Taylor}, {Thomas},
  {Tuffs}, {van Kampen}, \& {Wijesinghe}}]{Loveday2012}
{Loveday}, J., {Norberg}, P., {Baldry}, I.~K., {et~al.} 2012, \mnras, 420, 1239

\bibitem[{{Mao} {et~al.}(2021){Mao}, {Geha}, {Wechsler}, {Weiner}, {Tollerud},
  {Nadler}, \& {Kallivayalil}}]{Mao2021}
{Mao}, Y.-Y., {Geha}, M., {Wechsler}, R.~H., {et~al.} 2021, \apj, 907, 85

\bibitem[{{Mar{\'\i}n-Franch} \& {Aparicio}(2002)}]{Franch2002}
{Mar{\'\i}n-Franch}, A. \& {Aparicio}, A. 2002, \apj, 568, 174

\bibitem[{{Marshall} {et~al.}(1983){Marshall}, {Tananbaum}, {Avni}, \&
  {Zamorani}}]{Marshall1983}
{Marshall}, H.~L., {Tananbaum}, H., {Avni}, Y., \& {Zamorani}, G. 1983, \apj,
  269, 35

\bibitem[{{Marzke} {et~al.}(1998){Marzke}, {da Costa}, {Pellegrini}, {Willmer},
  \& {Geller}}]{Marzke1998}
{Marzke}, R.~O., {da Costa}, L.~N., {Pellegrini}, P.~S., {Willmer}, C. N.~A.,
  \& {Geller}, M.~J. 1998, \apj, 503, 617

\bibitem[{{Matthews} \& {Newman}(2010)}]{Matthews2010}
{Matthews}, D.~J. \& {Newman}, J.~A. 2010, The Astrophysical Journal, 721, 456

\bibitem[{{Matthews} \& {Newman}(2012)}]{Matthews2012}
{Matthews}, D.~J. \& {Newman}, J.~A. 2012, The Astrophysical Journal, 745, 180

\bibitem[{{McGaugh}(1996)}]{McGaugh1996}
{McGaugh}, S.~S. 1996, \mnras, 280, 337

\bibitem[{{McQuinn} \& {White}(2013)}]{McQuinn2013}
{McQuinn}, M. \& {White}, M. 2013, \mnras, 433, 2857

\bibitem[{{M{\'e}nard} {et~al.}(2013){M{\'e}nard}, {Scranton}, {Schmidt},
  {Morrison}, {Jeong}, {Budavari}, \& {Rahman}}]{Menard2013}
{M{\'e}nard}, B., {Scranton}, R., {Schmidt}, S., {et~al.} 2013, arXiv e-prints,
  arXiv:1303.4722

\bibitem[{{Mieske} {et~al.}(2012){Mieske}, {Hilker}, \& {Misgeld}}]{Mieske2012}
{Mieske}, S., {Hilker}, M., \& {Misgeld}, I. 2012, \aap, 537, A3

\bibitem[{{Moffett} {et~al.}(2016){Moffett}, {Lange}, {Driver}, {Robotham},
  {Kelvin}, {Alpaslan}, {Andrews}, {Bland-Hawthorn}, {Brough}, {Cluver},
  {Colless}, {Davies}, {Holwerda}, {Hopkins}, {Kafle}, {Liske}, \&
  {Meyer}}]{Moffett2016}
{Moffett}, A.~J., {Lange}, R., {Driver}, S.~P., {et~al.} 2016, \mnras, 462,
  4336

\bibitem[{{Newman}(2008)}]{Newman2008}
{Newman}, J.~A. 2008, The Astrophysical Journal, 684, 88

\bibitem[{{Peebles}(1980)}]{Peebles1980}
{Peebles}, P. J.~E. 1980, in Some Strangeness in the Proportion, ed.
  H.~{Woolf}, 302

\bibitem[{{Peebles} \& {Hauser}(1974)}]{Peebles1974}
{Peebles}, P.~J.~E. \& {Hauser}, M.~G. 1974, The Astrophysical Journal
  Supplement Series, 28, 19

\bibitem[{{Peng} {et~al.}(2010){Peng}, {Lilly}, {Kova{\v{c}}}, {Bolzonella},
  {Pozzetti}, {Renzini}, {Zamorani}, {Ilbert}, {Knobel}, {Iovino}, {Maier},
  {Cucciati}, {Tasca}, {Carollo}, {Silverman}, {Kampczyk}, {de Ravel},
  {Sanders}, {Scoville}, {Contini}, {Mainieri}, {Scodeggio}, {Kneib}, {Le
  F{\`e}vre}, {Bardelli}, {Bongiorno}, {Caputi}, {Coppa}, {de la Torre},
  {Franzetti}, {Garilli}, {Lamareille}, {Le Borgne}, {Le Brun}, {Mignoli},
  {Perez Montero}, {Pello}, {Ricciardelli}, {Tanaka}, {Tresse}, {Vergani},
  {Welikala}, {Zucca}, {Oesch}, {Abbas}, {Barnes}, {Bordoloi}, {Bottini},
  {Cappi}, {Cassata}, {Cimatti}, {Fumana}, {Hasinger}, {Koekemoer},
  {Leauthaud}, {Maccagni}, {Marinoni}, {McCracken}, {Memeo}, {Meneux}, {Nair},
  {Porciani}, {Presotto}, \& {Scaramella}}]{Peng2010b}
{Peng}, Y.-j., {Lilly}, S.~J., {Kova{\v{c}}}, K., {et~al.} 2010, \apj, 721, 193

\bibitem[{{Phillipps}(1985)}]{Phillipps1985}
{Phillipps}, S. 1985, Monthly Notices of the Royal Astronomical Society, 212,
  657

\bibitem[{{Phillipps} \& {Disney}(1986)}]{Phillipps1986}
{Phillipps}, S. \& {Disney}, M. 1986, \mnras, 221, 1039

\bibitem[{{Phillipps} \& {Shanks}(1987)}]{Phillipps1987}
{Phillipps}, S. \& {Shanks}, T. 1987, \mnras, 227, 115

\bibitem[{{Planck Collaboration} {et~al.}(2013){Planck Collaboration}, {Ade},
  {Aghanim}, {Arnaud}, {Ashdown}, {Atrio-Barandela}, {Aumont}, {Baccigalupi},
  {Balbi}, {Banday}, {Barreiro}, {Bartlett}, {Battaner}, {Benabed},
  {Beno{\^\i}t}, {Bernard}, {Bersanelli}, {Bonaldi}, {Bond}, {Borrill},
  {Bouchet}, {Burigana}, {Cabella}, {Cardoso}, {Catalano}, {Cay{\'o}n},
  {Chary}, {Chiang}, {Christensen}, {Clements}, {Colombo}, {Coulais}, {Crill},
  {Cuttaia}, {Danese}, {D'Arcangelo}, {Davis}, {de Bernardis}, {de Gasperis},
  {de Rosa}, {de Zotti}, {Delabrouille}, {Dickinson}, {Diego}, {Dobler},
  {Dole}, {Donzelli}, {Dor{\'e}}, {D{\"o}rl}, {Douspis}, {Dupac}, {Efstathiou},
  {En{\ss}lin}, {Eriksen}, {Finelli}, {Forni}, {Frailis}, {Franceschi},
  {Galeotta}, {Ganga}, {Giard}, {Giardino}, {Gonz{\'a}lez-Nuevo}, {G{\'o}rski},
  {Gratton}, {Gregorio}, {Gruppuso}, {Hansen}, {Harrison}, {Helou},
  {Henrot-Versill{\'e}}, {Hern{\'a}ndez-Monteagudo}, {Hildebrandt}, {Hivon},
  {Hobson}, {Holmes}, {Hornstrup}, {Hovest}, {Huffenberger}, {Jaffe},
  {Jagemann}, {Jewell}, {Jones}, {Juvela}, {Keih{\"a}nen}, {Knoche}, {Knox},
  {Kunz}, {Kurki-Suonio}, {Lagache}, {L{\"a}hteenm{\"a}ki}, {Lamarre},
  {Lasenby}, {Lawrence}, {Leach}, {Leonardi}, {Lilje}, {Linden-V{\o}rnle},
  {L{\'o}pez-Caniego}, {Lubin}, {Mac{\'\i}as-P{\'e}rez}, {Maffei}, {Maino},
  {Mandolesi}, {Maris}, {Marshall}, {Martin}, {Mart{\'\i}nez-Gonz{\'a}lez},
  {Masi}, {Massardi}, {Matarrese}, {Matthai}, {Mazzotta}, {Meinhold},
  {Melchiorri}, {Mendes}, {Mennella}, {Mitra}, {Moneti}, {Montier}, {Morgante},
  {Munshi}, {Murphy}, {Naselsky}, {Natoli}, {N{\o}rgaard-Nielsen}, {Noviello},
  {Novikov}, {Novikov}, {Osborne}, {Pajot}, {Paladini}, {Paoletti},
  {Partridge}, {Pearson}, {Perdereau}, {Perrotta}, {Piacentini}, {Piat},
  {Pierpaoli}, {Pietrobon}, {Plaszczynski}, {Pointecouteau}, {Polenta},
  {Ponthieu}, {Popa}, {Poutanen}, {Pratt}, {Prunet}, {Puget}, {Rachen},
  {Rebolo}, {Reinecke}, {Renault}, {Ricciardi}, {Riller}, {Ristorcelli},
  {Rocha}, {Rosset}, {Rubi{\~n}o-Mart{\'\i}n}, {Rusholme}, {Sandri}, {Savini},
  {Schaefer}, {Scott}, {Smoot}, {Spencer}, {Stivoli}, {Sudiwala}, {Suur-Uski},
  {Sygnet}, {Tauber}, {Terenzi}, {Toffolatti}, {Tomasi}, {Tristram},
  {T{\"u}rler}, {Umana}, {Valenziano}, {Van Tent}, {Vielva}, {Villa},
  {Vittorio}, {Wade}, {Wandelt}, {White}, {Yvon}, {Zacchei}, \&
  {Zonca}}]{Planck2013}
{Planck Collaboration}, {Ade}, P.~A.~R., {Aghanim}, N., {et~al.} 2013, \aap,
  554, A139

\bibitem[{{Popesso} {et~al.}(2005){Popesso}, {B{\"o}hringer}, {Romaniello}, \&
  {Voges}}]{Popesso2005}
{Popesso}, P., {B{\"o}hringer}, H., {Romaniello}, M., \& {Voges}, W. 2005,
  \aap, 433, 415

\bibitem[{{Rahman} {et~al.}(2016{\natexlab{a}}){Rahman}, {M{\'e}nard}, \&
  {Scranton}}]{Rahman2016a}
{Rahman}, M., {M{\'e}nard}, B., \& {Scranton}, R. 2016{\natexlab{a}}, Monthly
  Notices of the Royal Astronomical Society, 457, 3912

\bibitem[{{Rahman} {et~al.}(2015){Rahman}, {M{\'e}nard}, {Scranton}, {Schmidt},
  \& {Morrison}}]{Rahman2015}
{Rahman}, M., {M{\'e}nard}, B., {Scranton}, R., {Schmidt}, S.~J., \&
  {Morrison}, C.~B. 2015, Monthly Notices of the Royal Astronomical Society,
  447, 3500

\bibitem[{{Rahman} {et~al.}(2016{\natexlab{b}}){Rahman}, {Mendez},
  {M{\'e}nard}, {Scranton}, {Schmidt}, {Morrison}, \&
  {Budav{\'a}ri}}]{Rahman2016b}
{Rahman}, M., {Mendez}, A.~J., {M{\'e}nard}, B., {et~al.} 2016{\natexlab{b}},
  Monthly Notices of the Royal Astronomical Society, 460, 163

\bibitem[{{Robotham} {et~al.}(2018){Robotham}, {Davies}, {Driver}, {Koushan},
  {Taranu}, {Casura}, \& {Liske}}]{Robotham2018}
{Robotham}, A.~S.~G., {Davies}, L.~J.~M., {Driver}, S.~P., {et~al.} 2018,
  \mnras, 476, 3137

\bibitem[{{Sandage} {et~al.}(1985){Sandage}, {Binggeli}, \&
  {Tammann}}]{Sandage1985}
{Sandage}, A., {Binggeli}, B., \& {Tammann}, G.~A. 1985, \aj, 90, 1759

\bibitem[{{Schechter}(1976)}]{Schechter1976}
{Schechter}, P. 1976, \apj, 203, 297

\bibitem[{{Schmidt} {et~al.}(2013){Schmidt}, {M{\'e}nard}, {Scranton},
  {Morrison}, \& {McBride}}]{Schmidt2013}
{Schmidt}, S.~J., {M{\'e}nard}, B., {Scranton}, R., {Morrison}, C., \&
  {McBride}, C.~K. 2013, Monthly Notices of the Royal Astronomical Society,
  431, 3307

\bibitem[{{Schneider} {et~al.}(2006){Schneider}, {Knox}, {Zhan}, \&
  {Connolly}}]{Schneider2006}
{Schneider}, M., {Knox}, L., {Zhan}, H., \& {Connolly}, A. 2006, The
  Astrophysical Journal, 651, 14

\bibitem[{{Scottez} {et~al.}(2016){Scottez}, {Mellier}, {Granett}, {Moutard},
  {Kilbinger}, {Scodeggio}, {Garilli}, {Bolzonella}, {de la Torre}, {Guzzo},
  {Abbas}, {Adami}, {Arnouts}, {Bottini}, {Branchini}, {Cappi}, {Cucciati},
  {Davidzon}, {Fritz}, {Franzetti}, {Iovino}, {Krywult}, {Le Brun}, {Le
  F{\`e}vre}, {Maccagni}, {Ma{\l}ek}, {Marulli}, {Polletta}, {Pollo}, {Tasca},
  {Tojeiro}, {Vergani}, {Zanichelli}, {Bel}, {Coupon}, {De Lucia}, {Ilbert},
  {McCracken}, \& {Moscardini}}]{Scottez2016}
{Scottez}, V., {Mellier}, Y., {Granett}, B.~R., {et~al.} 2016, Monthly Notices
  of the Royal Astronomical Society, 462, 1683

\bibitem[{{Seldner} \& {Peebles}(1979)}]{Seldner1979}
{Seldner}, M. \& {Peebles}, P.~J.~E. 1979, \apj, 227, 30

\bibitem[{{Sinha} \& {Garrison}(2017)}]{corrfunc2017}
{Sinha}, M. \& {Garrison}, L. 2017, {Corrfunc: Blazing fast correlation
  functions on the CPU}, Astrophysics Source Code Library

\bibitem[{{Sprayberry} {et~al.}(1997){Sprayberry}, {Impey}, {Irwin}, \&
  {Bothun}}]{Sprayberry1997}
{Sprayberry}, D., {Impey}, C.~D., {Irwin}, M.~J., \& {Bothun}, G.~D. 1997,
  \apj, 482, 104

\bibitem[{{Swanson} {et~al.}(2008){Swanson}, {Tegmark}, {Blanton}, \&
  {Zehavi}}]{Swanson2008}
{Swanson}, M. E.~C., {Tegmark}, M., {Blanton}, M., \& {Zehavi}, I. 2008,
  \mnras, 385, 1635

\bibitem[{{Taylor} {et~al.}(2011){Taylor}, {Hopkins}, {Baldry}, {Brown},
  {Driver}, {Kelvin}, {Hill}, {Robotham}, {Bland-Hawthorn}, {Jones}, {Sharp},
  {Thomas}, {Liske}, {Loveday}, {Norberg}, {Peacock}, {Bamford}, {Brough},
  {Colless}, {Cameron}, {Conselice}, {Croom}, {Frenk}, {Gunawardhana},
  {Kuijken}, {Nichol}, {Parkinson}, {Phillipps}, {Pimbblet}, {Popescu},
  {Prescott}, {Sutherland}, {Tuffs}, {van Kampen}, \&
  {Wijesinghe}}]{Taylor2011}
{Taylor}, E.~N., {Hopkins}, A.~M., {Baldry}, I.~K., {et~al.} 2011, \mnras, 418,
  1587

\bibitem[{{Trentham} {et~al.}(2005){Trentham}, {Sampson}, \&
  {Banerji}}]{Trentham2005}
{Trentham}, N., {Sampson}, L., \& {Banerji}, M. 2005, \mnras, 357, 783

\bibitem[{{Trentham} \& {Tully}(2002)}]{Trentham2002}
{Trentham}, N. \& {Tully}, R.~B. 2002, \mnras, 335, 712

\bibitem[{{van Daalen} \& {White}(2018)}]{vanDaalen2018}
{van Daalen}, M.~P. \& {White}, M. 2018, Monthly Notices of the Royal
  Astronomical Society, 476, 4649

\bibitem[{{van den Busch} {et~al.}(2020){van den Busch}, {Hildebrandt},
  {Wright}, {Morrison}, {Blake}, {Joachimi}, {Erben}, {Heymans}, {Kuijken}, \&
  {Taylor}}]{Busch2020}
{van den Busch}, J.~L., {Hildebrandt}, H., {Wright}, A.~H., {et~al.} 2020,
  \aap, 642, A200

\bibitem[{{van Dokkum} {et~al.}(2015){van Dokkum}, {Abraham}, {Merritt},
  {Zhang}, {Geha}, \& {Conroy}}]{Dokkum2015}
{van Dokkum}, P.~G., {Abraham}, R., {Merritt}, A., {et~al.} 2015, \apjl, 798,
  L45

\bibitem[{{White} \& {Rees}(1978)}]{White1978}
{White}, S.~D.~M. \& {Rees}, M.~J. 1978, \mnras, 183, 341

\bibitem[{{Wright} {et~al.}(2017){Wright}, {Robotham}, {Driver}, {Alpaslan},
  {Andrews}, {Baldry}, {Bland-Hawthorn}, {Brough}, {Brown}, {Colless}, {da
  Cunha}, {Davies}, {Graham}, {Holwerda}, {Hopkins}, {Kafle}, {Kelvin},
  {Loveday}, {Maddox}, {Meyer}, {Moffett}, {Norberg}, {Phillipps}, {Rowlands},
  {Taylor}, {Wang}, \& {Wilkins}}]{Wright2017}
{Wright}, A.~H., {Robotham}, A.~S.~G., {Driver}, S.~P., {et~al.} 2017, \mnras,
  470, 283

\bibitem[{{Yamanoi} {et~al.}(2012){Yamanoi}, {Komiyama}, {Yagi}, {Okamura},
  {Iye}, {Kashikawa}, {Takata}, {Furusawa}, \& {Yoshida}}]{Yamanoi2012}
{Yamanoi}, H., {Komiyama}, Y., {Yagi}, M., {et~al.} 2012, \aj, 144, 40

\bibitem[{{Yamanoi} {et~al.}(2020){Yamanoi}, {Yagi}, {Komiyama}, \&
  {Koda}}]{Yamanoi2020}
{Yamanoi}, H., {Yagi}, M., {Komiyama}, Y., \& {Koda}, J. 2020, \aj, 160, 87

\bibitem[{{Zucca} {et~al.}(1997){Zucca}, {Zamorani}, {Vettolani}, {Cappi},
  {Merighi}, {Mignoli}, {Stirpe}, {MacGillivray}, {Collins}, {Balkowski},
  {Cayatte}, {Maurogordato}, {Proust}, {Chincarini}, {Guzzo}, {Maccagni},
  {Scaramella}, {Blanchard}, \& {Ramella}}]{Zucca1997}
{Zucca}, E., {Zamorani}, G., {Vettolani}, G., {et~al.} 1997, \aap, 326, 477

\end{thebibliography}

%%%%%%%%%%%%%%%%% APPENDICES %%%%%%%%%%%%%%%%%%%%
\onecolumn
\appendix
\section{Choice of model}
\label{app:choice of model}
\begin{figure}
	\centering
	\includegraphics[width=0.75\textwidth]{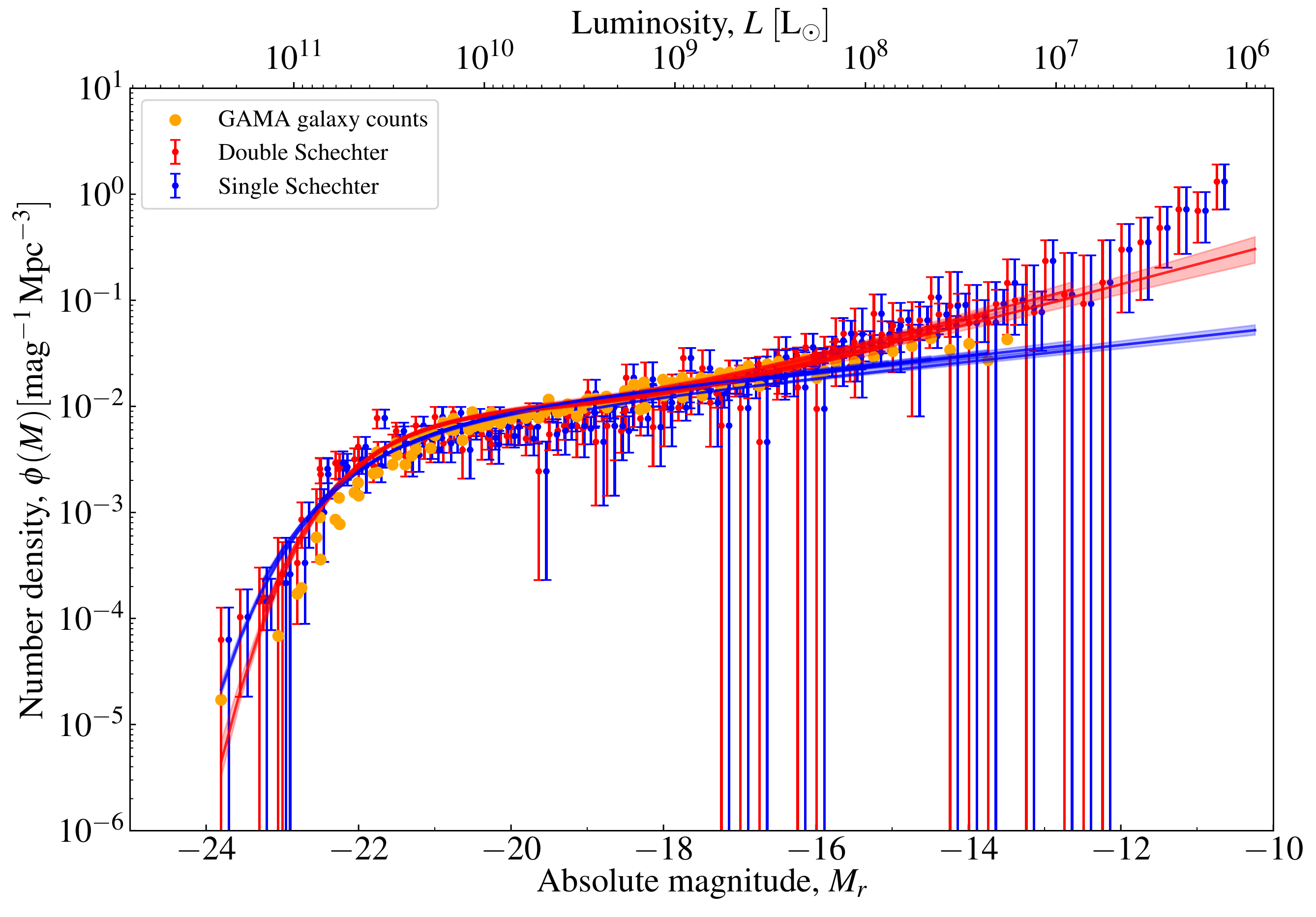}
	\caption{Density distribution of the luminosity function at $z<0.1$. Here the resulting model of the double Schechter fit (blue solid line) is compared to the single Schechter model (red solid line) as well as their corresponding A-factor normalised cluster-$z$s (blue/red line with error bars) and the measured GAMA values (orange solid points).}
	\label{fig:SinglevsDouble}
\end{figure}

The choice of parameterised GLF model is one possible limitation in our results, as the measurements depend somewhat on the parametrisation used. As we stated in the text we do not consider the best-fitting LF model as our results, but rather the inferred GLF measurements using the cluster-$z$s. In order to demonstrate that these measurements are insensitive to changes in the model parametrisation, we re-ran our analysis using a single Schechter function, rather than a double Schechter function, knowing that this is a poor choice of model. In Fig. \ref{fig:SinglevsDouble} the resulting A-factor normalised cluster-$z$s are displayed with their corresponding best-fitting models. It can be seen that the difference in the resulting values of the A-factor normalised cluster-$z$s is less than one standard deviation, while their corresponding models differ significantly. Hence we conclude that the results are robust to changes in the model parametrisation. In addition it can  be seen how the single Schechter formalism is a poor description of the data, as it is not able to describe the data well, and the double Schechter function model performs better at this task. This agrees with the literature results regarding the shape of the GLF.

\section{Sensitivity to stars and other objects}
\label{app:stellar_contribution}
\begin{figure}
	\centering
	\includegraphics[width=0.9\textwidth]{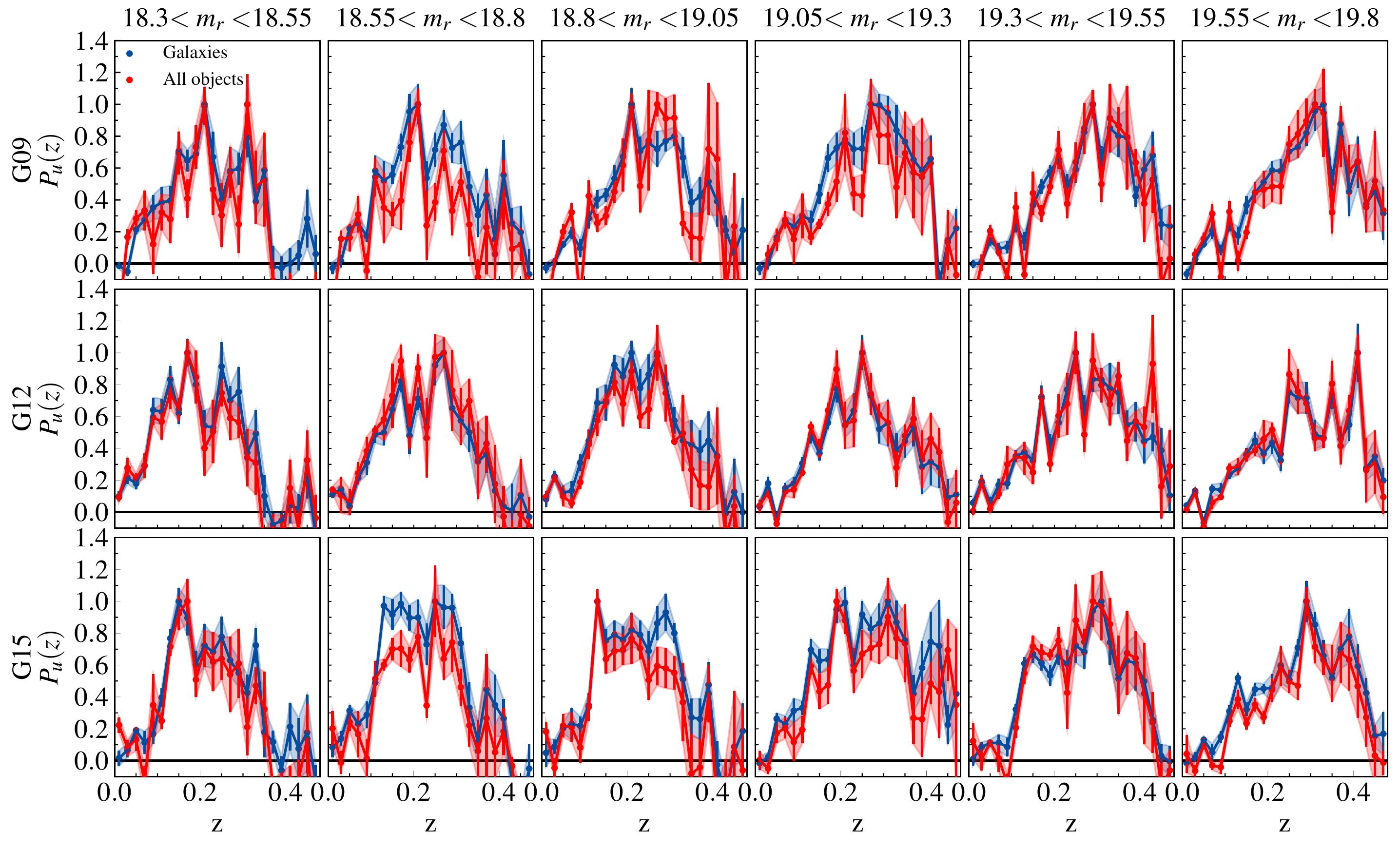}
	\caption{Comparison of the resulting $P_{m,z}$ using a target data set with mainly galaxies (blue) and one including all objects (red). The maximum value of all the $P_{m,z}$s are set to 1 for comparison.}
	\label{fig:StarGalaxySep}
\end{figure}

In Sec. \ref{sec:data} we explained how we excluded data flagged as stars within our target dataset. In order to demonstrate the effect of the stellar population on the resulting $P_{m,z}$, we recalculated the cluster-$z$s using all objects of the target dataset instead. By comparing the resulting $P_{m,z}$ with the original $P_{m,z}$, as can be seen in Fig. \ref{fig:StarGalaxySep}, the inclusion of the additional $ 30\% $ of data points, mainly consisting of stars, only has a  limited impact on the shape of the cluster-$z$s. Even such a large contamination of the data only produces a small impact on the results because, as mentioned in the text, the clustering amplitude only changes by a normalisation factor. For comparison we have normalised the resulting cluster-$z$s in Fig. \ref{fig:StarGalaxySep} such that their maximum equals 1. The different amplitude is of no concern, as the $A_m$ factor accounts for any global changes of the amplitude. We can hence conclude that our technique is insensitive to stellar contamination.

%%%%%%%%%%%%%%%%%%%%%%%%%%%%%%%%%%%%%%%%%%%%%

% Don't change these lines
%\bsp   % typesetting comment
\label{lastpage}
\end{document}